# Few-Shot Open-Set Audio Classification Using Attention Information-Fused Prototypes

Yanxiong Li, *Senior Member*, *IEEE*, Jiaxin Tan, Qianqian Li, Guoqing Chen, Sen Huang, Tuomas Virtanen, *Fellow*, *IEEE*

***Abstract*—Most existing audio classification methods suppose that each query (testing) sample belongs to a class of support (training) samples, and misrecognize samples of unseen classes as seen classes (cannot reject samples of unseen classes). In this study, we propose a method for Few-shot Open-set Audio Classification (FOAC), which can recognize query samples of seen classes after updating the model using a few support samples, and meanwhile reject query samples from unseen classes. We design a model consisting of an encoder and a classifier. The encoder is the backbone of a ResNet used for extracting embeddings. The classifier consists of prototype generators of few-shot classes and open-set classes. Prototypes of few-shot classes are obtained by fusing the class-discriminative information of support and query embeddings and by assigning larger weighting coefficient to representative part of the support embeddings. One prototype is generated for open-set classes using the proposed prototype generator. The encoder is trained with abundant samples of base classes in supervised manner, and then the prototypes of base classes are generated under the supervision of a joint loss. The classifier is trained using a few samples of few-shot classes in a meta-training way. Three public datasets (LS-100, NSynth-100, and FSC-89) are used to assess the performance of our method. Experiments show that our method has advantage over prior methods in AUROC and accuracy. This advantage has statistical significance for most prior methods. Our method has lower computational complexity than most prior methods. The code is at https://github.com/Jessytan/FOAC-AIFP.**

## I. Introduction

AUDIO classification (AC) aims to classify various types of audios, which is a key module for realizing audio or video related processing tasks, such as sound event detection [1]-[4], acoustic scene classification [5]-[8], video content analysis [9], [10], speaker diarization or recognition [11]-[14], keyword spotting [15]-[18]. Realistic soundscapes are dynamic and open. New audio classes (such as the sound of new car engines) are constantly appearing after initially training models, yet labeled samples for these new audio classes are generally scarce. This has given rise to Few-shot Open-set AC (FOAC). The FOAC requires models to learn the features of new classes using a few labeled samples (few-shot), and also to distinguish between unseen classes and seen classes (open-set). The motivation for FOAC stems from the needs of practical applications. First, data annotation is expensive. Accurate manual annotation of massive data is time-consuming and laborious. Second, real-world acoustic environments are inherently open and dynamic. Whether it is the sound of new devices in smart homes or the discovery of new species in bioacoustic monitoring, models should be able to continuously adapt and learn new audio classes, rather than being restricted by a predefined fixed set of audio classes.

This work was partly supported by the national natural science foundation of China (62371195, 62111530145), exchange project of the 10th Meeting of China-Croatia Science and Technology Cooperation Committee (10-34).
*(Corresponding author: Yanxiong Li,* eeyxli@scut.edu.cn*.)*

Yanxiong Li, Jiaxin Tan, Qianqian Li, Guoqing Chen and Sen Huang are with the School of Electronic and Information Engineering, South China University of Technology, Guangzhou 510640, China.

Tuomas Virtanen is with the Faculty of Information Technology and Communication Sciences, Tampere University, FI-33720 Tampere, Finland.
**None of the authors have a conflict of interest to disclose**.

Traditional supervised learning methods rely on large-scale, fixed-category annotated datasets for training. When faced with new audio classes, it is often required to collect a large number of new samples and retrain the model from scratch, which is neither economical nor feasible in practice. Some prior methods are effective for addressing two main issues of the FOAC [19]-[23], namely recognizing novel classes after updating the model with a few support samples and rejecting unseen classes (not recognizing unseen classes as seen classes). However, they have the following two limitations.

First, they focus on learning discriminative embeddings from the samples of seen classes, without considering the representations of unseen classes. Focusing only on seen classes will result in significant misclassification risks for unseen classes since information about unseen classes is not considered at all. Effective FOAC solutions should optimize the processing for unseen classes and seen classes. How to effectively characterize unseen classes using only training samples from seen classes is a challenging problem and will be discussed in this work.

Second, the prototypes of base classes and novel classes (defined in Fig. 1) are generated without fusing multiple types of information, where a prototype is a representative vector of embeddings for a given class. For example, the prototypes of few-shot classes are constructed without fusing the information of support and query samples. Moreover, the prototypes of open-set classes are not considered or are generated without using the information of prototypes of base classes. In addition, all parts (dimensions) of an embedding are treated in the same way when they are used to generate prototypes. Representative parts of the embedding are overwhelmed by useless parts.

To overcome the above two weaknesses of current methods, we propose a FOAC method using attention information-fused prototypes. To facilitate readers' understanding of various audio classes, Fig. 1 lists their definitions. The work in this paper has the following three contributions.

1. We propose to generate the prototypes of base classes

Audio classes
- Base Classes (BC): has abundant samples to pre-train models, whose prototypes are generated by their samples.
- Augmented Base-classes (AB): has no samples, whose prototypes are generated by the BC's samples.
- Pseudo Novel-classes (PN): used to simulate novel-classes for meta-training models with support and query samples, whose samples are from BC.
- Novel classes: used to meta-train models with few samples and to test models.
  - Few-shot Classes (FC): has support and query samples, whose prototypes are generated by their support samples.
  - Open-set Classes (OC): has query samples only, whose prototypes are generated by prototypes of BC, AB and FC.

Fig. 1. The divisions (black text) and definitions (blue text) of audio classes that disjoint from each other and are used for model training or testing.

and augmented base-classes for FOAC, which is not considered in previous work. When abundant samples of base classes are adopted to generate the prototypes of base classes with supervised training, the prototype of augmented base-class corresponding to a base class is forced to move away from the embeddings extracted from the samples of this base class. Hence, the generated prototypes of augmented base-classes are expected to be far away from prototypes (representative vectors of embeddings) of base classes.

2. We design a model which consists of an encoder with a common structure and a classifier with a new structure. The encoder is a convolutional neural network, while the classifier consists of a Prototype Generator of Few-shot Classes (PGFC) and a Prototype Generator of Open-set Classes (PGOC). The PGFC generates prototypes of few-shot classes by fusing the class-discriminative information of both support and query embeddings, and by assigning larger weighting coefficient to representative part of the support embeddings transformed by the PGFC. The PGOC generates a prototype for open-set classes by fusing information of the prototypes of few-shot classes, base classes and augmented base-classes. In short, the prototypes of few-shot and open-set classes are generated by fusing multiple information using the attention modules.

3. We propose a FOAC method using attention information-fused prototypes to recognize query samples from few-shot classes and reject query samples from open-set classes. The attention information-fused prototypes are generated by the PGFC and PGOC, and can construct a discriminative decision boundary for distinguishing between the few-shot classes and the open-set classes. Our method's effectiveness is validated from multiple views on three public datasets (LS-100, FSC-89 and NSynth-100). Results show that it outperforms previous methods in AUROC and Accuracy, and that it has lower computational complexity than most of previous methods.

## II. Related Work

This section describes related work about few-shot learning, few-shot open-set recognition and AC.

### A. *Few-shot Learning*

Few-shot learning is a problem in which the models aim to adapt and can generalize (not overfit to) given a few training samples. Main solutions include the methods of meta-learning, transfer learning, and data augmentation.

Meta-learning methods are the dominant solution to address the few-shot learning problem, including the metric-based, model-based, optimization-based, probability-based, and neural-processes-based methods [24]. The metric-based methods aim to learn an embedding space where it is easier to classify the embeddings of different classes [25], [26]. The model-based methods aim to develop specific models that enable rapid adaptation to new tasks. They are based on either architecture or memory [27], [28]. The goal of optimization-based methods is to learn the processes that are suitable for optimization across various tasks without multiple iterations or massive data. They mainly include learning the update rule [29], learning the initial parameters [30], and learning to control the update rule [31]. To overcome the limitations of deterministic meta-learning methods, probability-based methods are proposed to improve uncertainty quantification and performance via modeling parameters as random variables and generating their distributions [32]. To extend the meta-learning paradigm, many efforts are made on the neural-processes-based methods [33].

Transfer learning methods generally rely on a two-stage learning process that includes pre-training a feature extractor on a large-scale base dataset and executing finetuning in downstream new tasks with a few samples [34]. They are more lightweight compared to the meta-learning methods.

Data augmentation can diversify training data, thereby solving the problem of insufficient data in few-shot learning. This is main function of data augmentation methods, which reduces overfitting by enforcing regularization with the existing data or by increasing the number of samples [35].

### B. *Few-shot Open-set Recognition*

Few-shot open-set recognition is an under-explored problem, in which open-set recognition is combined with few-shot learning to solve open-set recognition issues with limited training samples. Open-set recognition is a problem in which a model must correctly recognize testing samples from seen classes appeared in model training and not recognize testing samples belonging to unseen classes as seen classes. We introduce some typical few-shot open-set recognition methods in next paragraph of this subsection.

Huang et al. [36] propose a few-shot open-set recognition method based on Task-Adaptive Negative Envision (TANE), in which the TANE is used to integrate decision-threshold-tuning into the learning process. The few-shot closed-set classifier is augmented by additional negative prototypes that are generated with few-shot examples. In the method of Few-shot Embedding Adaptation with Transformer (FEAT) [37], embeddings are adapted to the target classification task with a set-to-set function, yielding embeddings that are task-specific and discriminative. Different instantiations of such set-to-set functions are investigated and the Transformer is found to be the most effective. Wang et al. [38] design a model of Glocal Energy-based Learning (GEL) to solve the few-shot open-set recognition problem. The GEL model consists of two branches, where a classification branch is used to learn a metric for recognizing a sample as one of seen classes and an energy branch is adopted to estimate the probability of unseen classes. Sun et al. [39] propose a method by using Overall Positive Prototype (OPP) where an OPP is generated to represent positive samples. Each query sample is recognized as positive or negative through measuring the distance between it and overall positive prototype. Sapkota et al. [40] propose a method using Meta Evidential Transformer (MET), where a MET-based model is designed to learn the compact embeddings of closed-set classes by similar closed-set classes.

### C. Audio Classification

According to the model's abilities to recognize novel classes and reject unseen classes, and number of training samples, existing AC methods are split into five types, including Many-shot AC (MAC), Few-shot AC (FAC), Class-incremental AC (CAC), Few-shot Class-incremental AC (FCAC), and FOAC.

In the many-shot AC methods, the models are typically trained in a supervised way with enough samples per class [41]-[44]. The identity of classes needs to be predefined in advance and the number of classes is fixed. The model can recognize the predefined classes, but cannot recognize novel classes. The model must be retrained using sufficient training data of predefined and novel classes to recognize their testing data. It is troublesome and even infeasible for ordinary users to retrain the model due to the following two factors. First, due to data copyright protection and privacy concerns, the training data of predefined classes are often not accessed for ordinary users after training the model. Second, ordinary users usually lack the skills and powerful computers to retrain the models. If a model is retrained with the samples of novel classes only, the knowledge of predefined classes will be forgotten. The many-shot AC methods can recognize the predefined classes after training the model with abundant samples, but cannot recognize novel classes and reject unseen classes.

To reduce the requirement for the number of training samples, the few-shot AC methods are proposed [45]-[48]. They recognize novel classes using a few instead of abundant training samples per novel class. This ability is their most significant advantage over the many-shot AC methods. To enable the model to recognize novel classes, an encoder which can extract discriminative and generalizable embeddings must be trained with sufficient training samples from base classes in advance. Although the few-shot AC methods can recognize novel classes using a few training samples of novel classes, they are prone to overfitting to novel classes. Besides, they cannot memorize old-classes and reject unseen classes.

The class-incremental AC methods can continually recognize novel classes and memorize old-classes [49]-[51]. They can recognize novel classes after updating the model with sufficient samples from each novel class and memorize old-classes in all sessions. However, abundant training samples of old-classes and novel classes are required to train (update) model in different stages. Like the many-shot AC methods, the class-incremental AC methods also have the problem of high cost in collecting many training samples. Some audio classes have relatively scarce samples, such as endangered animals' call. It is difficult to collect sufficient samples of these audio classes. Hence, the class-incremental AC methods have limitations in applications involving the audio classes with scarce samples.

The few-shot class-incremental AC methods can continually recognize novel classes and memorize old-classes using a few training samples from novel classes [52]-[57]. Moreover, a fully few-shot class-incremental AC method is proposed in [58], [59], which incrementally recognizes novel classes and memorize old-classes with a few training samples of base classes and novel classes. There are two types of sessions in the few-shot class-incremental AC methods, including base and incremental sessions. The model is decomposed into an encoder and a classifier. The encoder is frozen after training in the base session for remembering the knowledge of base classes. The classifier is updated in each incremental session to recognize novel classes. Main part for encoders is the backbone of a convolutional neural network [53]-[57]. Main difference between various methods is the classifiers. Main classifiers include stochastic classifier [52], [53], prototype classifier [54]-[56], and fully-connected-layer classifier [57]-[59]. The few-shot class-incremental AC methods can incrementally recognize novel classes and remember old ones, but they cannot reject query samples from unseen classes.

TABLE I
CHARACTERS OF VARIOUS TYPES OF AUDIO CLASSIFICATION METHODS

| Types | Training samples per | | Recognize base classes | Recognize novel classes | Reject unseen classes |
|---|---|---|---|---|---|
| | base class | novel class | | | |
| MAC | Abundant | × | √ | × | × |
| FAC | Abundant | Few | √ | √ | × |
| CAC | Abundant | Abundant | √ | √ | × |
| FCAC | Abundant | Few | √ | √ | × |
| FOAC | Abundant | Few | √ | √ | √ |

From the above descriptions, it is known that the methods of many-shot AC, few-shot AC, class-incremental AC and few-shot class-incremental AC suppose a pre-defined set of classes with labeled training samples. Each testing sample is assumed to belong to one of the pre-defined classes and thus these four types of methods only conduct a closed-set decision. In practical applications, we face more challenging situations. First, collecting adequate labeled training samples are difficult to guarantee due to high cost for collecting labeled training samples and the sensitivity or scarcity of samples from some classes. Second, many testing samples do not belong to the seen classes (classes of training samples). The above four types of methods will recognize the out-of-set testing samples as seen classes instead of rejecting them as unseen classes. Hence, they can recognize testing samples from seen classes but cannot reject testing samples from unseen classes.

To overcome the shortcoming of the above work that cannot reject unseen classes, some efforts are made for audio processing [19]-[23]. For example, Li et al. use multi-stage training [19] for few-shot open-set keyword spotting. Rusci et al. explore few-shot open-set keyword recognition using a feature encoder and a prototype classifier [20]. The Dummy Prototypical Network (D-ProtoNet) is used to generate episode known dummy prototypes with metric learning, and exceeds other keyword spotting methods [21]. In the OpenFEAT method [22], an embedding adaptation framework is used to adapt embeddings from the given universal embedding space to a specific embedding space by the Transformer-based set-to-set function. In the $L^3$-Net method [23], the encoder is trained by self-supervised learning of audio-visual correspondence in videos without requiring labeled data. The above work enables the model to reject unseen classes in the context of few-shot training, and can be regarded as FOAC methods. The characters of various methods are listed in Table I.

## III. METHOD

This section introduces our method in detail, including problem definition, method framework, prototype generation of augmented base-class, PGFC, and PGOC.

### A. Problem Definition

To let readers understand symbols in the paper, Table II

TABLE II
DEFINITIONS OF SYMBOLS USED IN THE PAPER

| Sym. | Definitions | Sym. | Definitions |
|---|---|---|---|
| $\boldsymbol{x}^s$ | Support sample of few-shot class | $\boldsymbol{x}^q$ | Query sample |
| $\boldsymbol{D}_t^{bc}$ | Training set of base classes | $\boldsymbol{e}^q$ | Embedding of $\boldsymbol{x}^q$ |
| $\boldsymbol{D}_t^{pn}$ | Training set of pseudo novel-classes | $\boldsymbol{e}^s$ | Embedding of $\boldsymbol{x}^s$ |
| $\boldsymbol{D}_q^{oc}$ | Query set of open-set classes | $\boldsymbol{D}_q$ | Query set of novel-classes |
| $\boldsymbol{L}^{fc}$ | Label set of few-shot classes | $\boldsymbol{\theta}$ | Encoder |
| $\boldsymbol{W}_q/\boldsymbol{W}_s$ | Weighting coefficients vector | $N$ | No. of few-shot classes |
| $\mathcal{L}^c$ | Loss of empirical classification risk | $\alpha$ | Weighting coefficient |
| $\mathcal{L}^s$ | Loss of complementary space risk | $\boldsymbol{M}$ | Vector of learnable margin |
| $M_h$ | Learnable margin for prototype $\boldsymbol{p}_h^{ab}$ | $\boldsymbol{P}'^{fc}$ | Negative prototype of $\boldsymbol{P}^{fc}$ |
| $\hat{\boldsymbol{e}}^q/\hat{\boldsymbol{e}}^s$ | Transformed query/support embedding | $\bar{\boldsymbol{P}}^{fc}$ | Mean prototype of $\boldsymbol{P}^{fc}$ |
| $D$ | Dimension of prototype or embedding | T | Transpose of matrix |
| $\boldsymbol{D}_s^{fc}/\boldsymbol{D}_q^{fc}$ | Support/Query set of few-shot classes | | |
| $\boldsymbol{D}_{s,i}^{fc}/\boldsymbol{D}_{q,i}^{fc}$ | Support/Query set of the $i$th few-shot class | | |
| $K$ | Number of support samples per few-shot class | | |
| $\boldsymbol{p}_i^{fc}$ | Prototype of the $i$th few-shot class | | |
| $\boldsymbol{p}_h^{bc}/\boldsymbol{p}_h^{ab}$ | Prototype of the $h$th base class/augmented base-class | | |
| $H$ | Number of base classes or augmented base-classes | | |
| $\boldsymbol{P}^{bc}/\boldsymbol{P}^{ab}$ | Prototypes of base classes/augmented base-classes | | |
| $\boldsymbol{P}^{fc}/\boldsymbol{p}^{oc}$ | Prototype of few-shot/open-set classes | | |
| $\boldsymbol{e}_{h,j}$ | Embedding of the $j$th sample of the $h$th base class | | |

lists their definitions. As shown in Fig. 2, the FOAC task has two sessions: model training and testing. The model training session includes pre-training and meta-training. In the pre-training step, the base classes' training set $\boldsymbol{D}_t^{bc}$ (a set of labeled audio samples) is used to pre-train a model which extracts discriminative embeddings and generalizes well for novel classes. The model is trained in a supervised manner by abundant samples per base class. In the meta-training step, the pseudo novel-classes' training set $\boldsymbol{D}_t^{pn}$ (a set of labeled audio samples) is used to meta-train the model. That is, the model is trained in a few-shot learning way with a few samples per pseudo novel-class in each episode. In the model testing step, support samples of few-shot classes' support set $\boldsymbol{D}_s^{fc}$ are used to meta-train the model. The trained model is used to recognize each query sample from few-shot classes' query set $\boldsymbol{D}_q^{fc}$ and reject each query sample from open-set classes' query set $\boldsymbol{D}_q^{oc}$.

A FOAC task for model testing is denoted as $\Gamma = (\boldsymbol{D}_s^{fc}, \boldsymbol{D}_q^{fc}, \boldsymbol{D}_q^{oc}, \boldsymbol{L}^{fc})$ which means that after updating the model with $\boldsymbol{D}_s^{fc}$ and $\boldsymbol{L}^{fc}$ (label set of few-shot classes), the model can recognize the samples in $\boldsymbol{D}_q^{fc}$ and classify the samples in $\boldsymbol{D}_q^{oc}$ as unseen classes. In addition, $\boldsymbol{D}_s^{fc} = \{\boldsymbol{D}_{s,i}^{fc}, 1 \le i \le N\}$, $\boldsymbol{D}_q^{fc}=\{\boldsymbol{D}_{q,i}^{fc}, 1 \le i \le N\}$, $|\boldsymbol{L}^{fc}|=N$, $|\boldsymbol{D}_{s,i}^{fc}|=K$, and $\boldsymbol{D}_q=\boldsymbol{D}_q^{fc} \cup \boldsymbol{D}_q^{oc}$, and $|\cdot|$ denotes the cardinality. $\boldsymbol{D}_{s,i}^{fc}$ and $\boldsymbol{D}_{q,i}^{fc}$ are support and query set of the $i$th few-shot class, respectively. $N$, $K$ and $\boldsymbol{D}_q$ are the number of few-shot classes, number of support samples per few-shot class and query set of novel classes, respectively. The problem based on the above setting is called a $N$-way $K$-shot FOAC task.

Prototypical network [26] is utilized to learn a prototype-based classifier. The $i$th few-shot class can be represented by a prototype $\boldsymbol{p}_i^{fc}$ which can be obtained by calculating the mean vector of embeddings of $K$ support samples of the $i$th few-shot class. Namely, $\boldsymbol{p}_i^{fc}$ is calculated by

$$\boldsymbol{p}_i^{fc} = \frac{1}{K}\sum_{\boldsymbol{x}^s \in \boldsymbol{D}_{s,i}^{fc}} \boldsymbol{e}^s , \tag{1}$$

where $\boldsymbol{x}^s$ denote a support sample of few-shot class, and $\boldsymbol{e}^s$

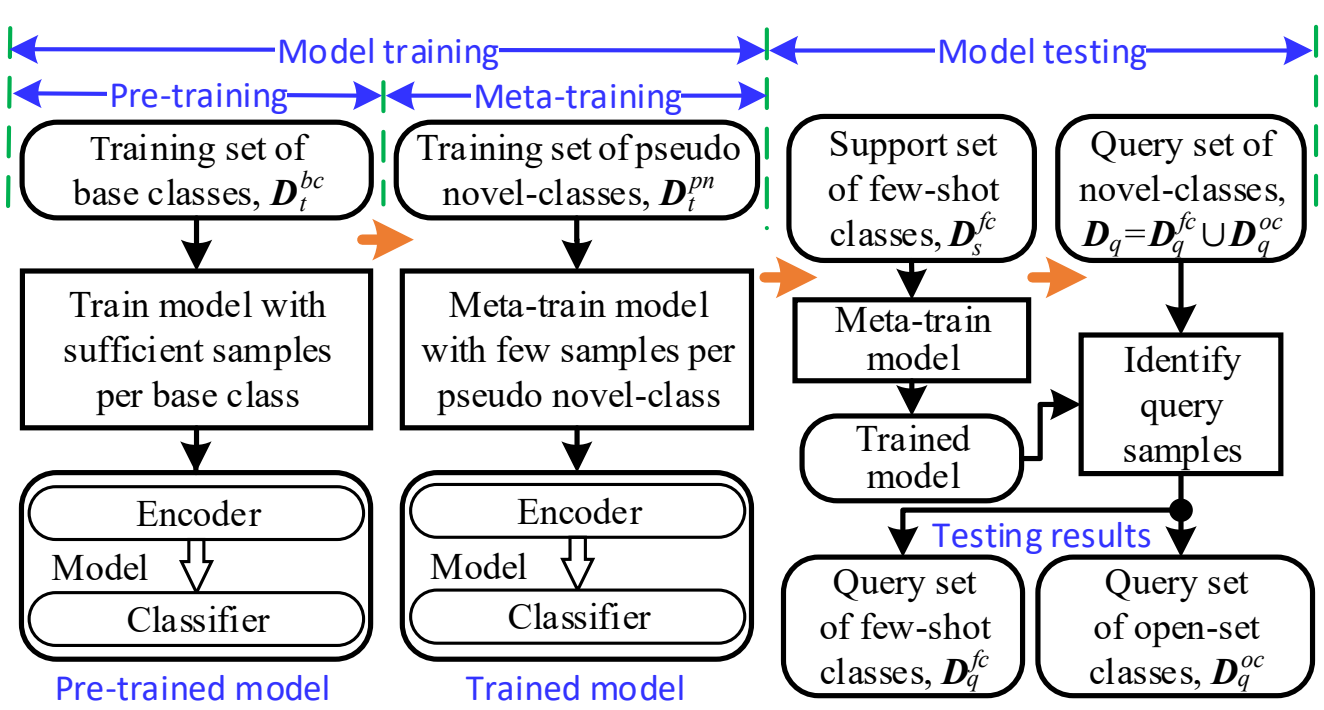

Fig. 2. The definition of few-shot open-set audio classification. Orange arrows represent the transition between different stages.

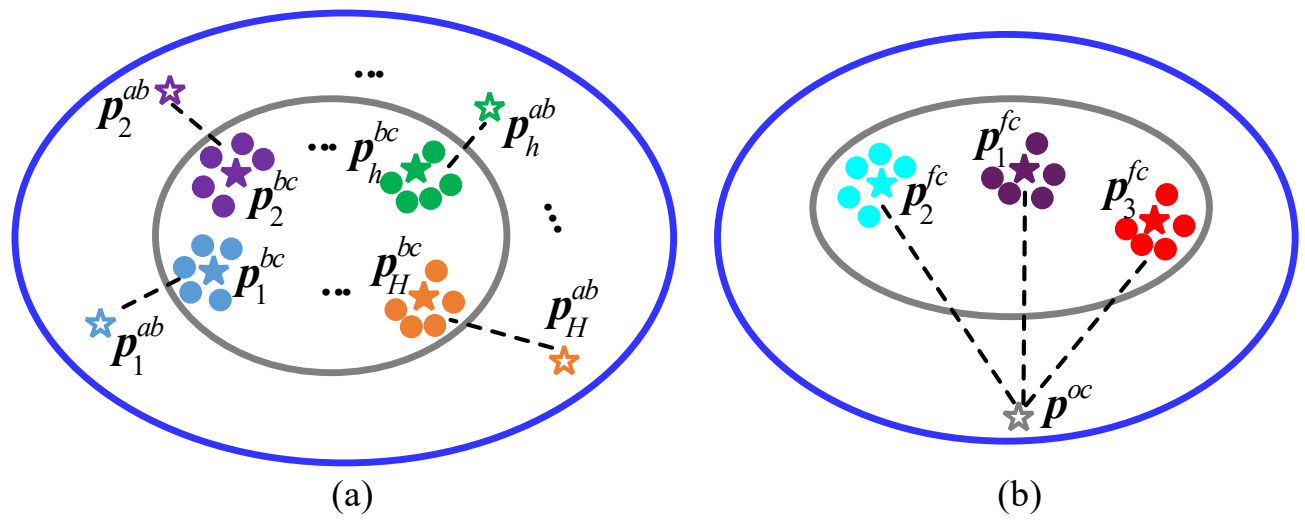

Fig. 3. (a) Illustrations of $H$ base classes' prototypes (solid pentagrams) and embeddings (solid dots), and $H$ augmented base-classes' prototypes (hollow pentagrams), where $1 \le h \le H$; (b) illustrations of three few-shot classes' prototypes (solid pentagrams) and embeddings (solid dots), and one open-set class's prototypes (hollow pentagrams).

represents embedding extracted by the encoder from $\boldsymbol{x}^s$. All prototypes $\boldsymbol{P}^{fc} = \{\boldsymbol{p}_i^{fc}\}$ form a closed-set classifier where a query sample $\boldsymbol{x}^q \in \boldsymbol{D}_q^{fc}$ is recognized by a nearest neighbor search, i.e.

$$argmin_i \left(\{f_d(\boldsymbol{e}^q, \boldsymbol{p}_i^{fc})\}_{i \in \boldsymbol{L}^{fc}}\right), \tag{2}$$

where $\boldsymbol{e}^q$ and $f_d(\cdot,\cdot)$ denote embedding of query sample $\boldsymbol{x}^q$ and a function for measuring the distance between two inputs (e.g., cosine distance), respectively.

To recognize a query sample as a few-shot class or reject it as open-set class, the closed-set classifier needs to be calibrated as an open-set classifier and a threshold $\xi$ needs to be set. A query sample $\boldsymbol{x}^q$ in open-set classes is rejected if the distances between $\boldsymbol{e}^q$ and prototypes of all few-shot classes exceed $\xi$, i.e., $min\left(\{f_d(\boldsymbol{e}^q, \boldsymbol{p}_i^{fc})\}_{i \in \boldsymbol{L}^{fc}}\right) > \xi$. To adaptively reject the query samples of open-set classes, we do not set a fixed threshold, but generate a prototype for the open-set classes.

## *B. Method Framework*

Because of the large number of base classes and abundant samples of each base class in the pre-training step, we construct an augmented base-class for each base class. Augmented base-classes do not have samples, and their prototypes are generated by the samples of base classes (detailed in subsection III-C). Fig. 3 (a) illustrates the prototypes of $H$ base classes ($\boldsymbol{p}_1^{bc}$ to $\boldsymbol{p}_H^{bc}$, marked by solid pentagrams) and $H$ augmented base-classes ($\boldsymbol{p}_1^{ab}$ to $\boldsymbol{p}_H^{ab}$, marked by hollow pentagrams). The area within the gray elliptical curve is the embedding space represented by the prototypes of base classes, while the area within the blue elliptical curve is the embedding space represented by the prototypes of both base classes and augmented base-classes. As illustrated in Fig. 3 (a), the prototypes of both base classes and augmented base-classes represent a larger embedding space and thus are expected to provide more useful cues for generating the prototypes of open-set classes.

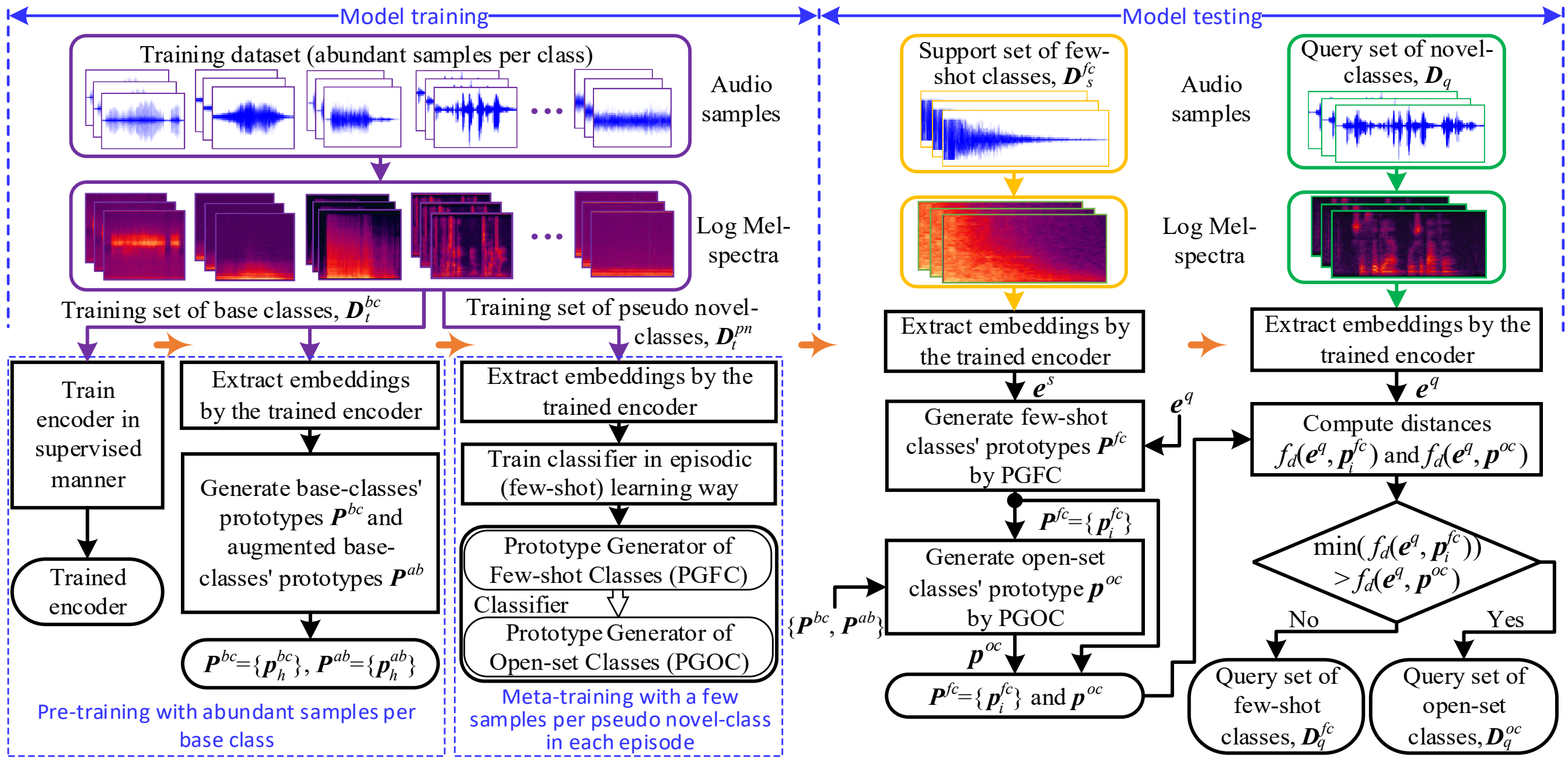

Fig. 4. Framework for the proposed FOAC method. Orange arrows represent the transition between different stages. $\boldsymbol{e}^s$: embedding of a support sample of novel-classes; $\boldsymbol{e}^q$: embedding of a query sample of novel-classes; $\boldsymbol{p}_i^{fc}$: prototype of the $i$th few-shot class; $\boldsymbol{p}^{oc}$: prototype of open-set classes; $\min(f_d(\boldsymbol{e}^q, \boldsymbol{p}_i^{fc}))$: minimum of distances between $\boldsymbol{e}^q$ and $\boldsymbol{p}_i^{fc}$. Here, $1 \le h \le H$ and $1 \le i \le N$. $H$ denotes the number of base classes (or augmented base-classes), while $N$ represents the number of few-shot classes.

To effectively distinguish between few-shot and open-set classes, we need to construct a prototype for open-set classes. Each few-shot class only has a few samples, which is not sufficient to construct a prototype with strong representational ability for each open-set class. Hence, we construct only one prototype of open-set classes for all few-shot classes. The open-set class does not have support samples, and its prototype is generated by the prototypes of few-shot classes, base classes and augmented base-classes. Fig. 3 (b) depicts the prototypes of three few-shot classes and one open-set class.

The framework of the proposed method is shown in Fig. 4. The model consists of an encoder and a classifier. The samples of unseen and seen classes pass through the same trained encoder. The framework has a model training session (including pre-training and meta-training steps) and a model testing session. In the model training session, the training set of base classes (i.e., $\boldsymbol{D}_t^{bc}$) and the training set of pseudo novel-classes (i.e., $\boldsymbol{D}_t^{pn}$) are from the training dataset. In the pre-training step, $\boldsymbol{D}_t^{bc}$ is used to train encoder by supervised learning. Then, the prototypes of both base classes and augmented base-classes are generated. In the meta-training step, $\boldsymbol{D}_t^{pn}$ is used to train the classifier in a way of episodic (few-shot) learning using a cross-entropy loss. In the model testing session, the prototypes of few-shot classes are generated by support and query embeddings, and the prototypes of open-set classes are generated by the prototypes of few-shot classes, base classes and augmented base-classes. Then, each query sample is recognized as a seen class or rejected as unseen classes based on the distances between its embedding and prototypes of all novel classes. The reasons for taking the above measures are as follows.

First, to overcome the collection difficulty of pseudo novel-classes' samples for model training and to obtain a classifier for generating representative prototypes of novel classes, the training dataset is used to train both the encoder in the pre-training step and the classifier in the meta-training step. Pseudo novel classes are the assumed novel classes and thus can be created from any dataset except for those containing true novel classes. We can obtain a lot of base classes and abundant samples per base class for training the model in practice. Therefore, we divide the training dataset into a large-scale dataset with a larger number of classes and samples (i.e., $\boldsymbol{D}_t^{bc}$) and a small-scale dataset with a few classes and samples (i.e., $\boldsymbol{D}_t^{pn}$). The classes and samples in $\boldsymbol{D}_t^{bc}$ and $\boldsymbol{D}_t^{pn}$ are different. $\boldsymbol{D}_t^{pn}$ is constructed by randomly selecting classes and the classes' samples from the training dataset.

Second, the encoder and classifier have different functions in the FOAC method and thus need to be trained in different ways. The encoder is used to extract embeddings from samples in all steps. Hence, the encoder should be generated first, and be trained using a lot of samples from each base class to ensure that embeddings extracted by the encoder have strong discriminative and generalization abilities for all classes. The classifier consists of a PGFC and a PGOC, which is utilized to generate prototypes of few-shot classes and open-set classes. In the FOAC task, each few-shot class has only a few support (training) samples. To ensure the trained classifier to perform well in the testing session (with only a few support samples per class), the training way of the classifier should match the testing way. Namely, the classifier should be trained in a way of $N$-way $K$-shot episodic learning. Hence, a few samples per pseudo novel-class are used to train the classifier in each episode of the meta-training step.

Third, prototypes of base classes and augmented base-classes benefit for enhancing the representational ability of prototypes of novel classes. The prototypes of base classes and augmented base-classes are generated by rich training data of base classes and are expected to have strong representational ability. If the prototypes of base classes and augmented base-classes are used to generate the novel classes' prototypes, the representational ability of the prototypes of novel classes can be enhanced. We use sufficient training data to generate prototypes of base classes and augmented base-classes in the pre-training step.

The feature of log Mel-spectrum [60] is extracted from each sample in all steps for model training and testing and fed to the encoder to extract embeddings. It can represent time-frequency properties of samples well and is widely used as the input of neural network to learn embeddings [61], [62].

### *C. Prototype Generation of Augmented Base-class*

This subsection describes prototype generation of base classes and augmented base-classes. The trained encoder is used to extract embeddings. The prototypes of base classes and augmented base-classes are used to represent their respective embedding space and to generate prototypes of novel classes.

The ResNet-based network [63] has powerful capability to extract discriminative embeddings from samples, which has been proved by its wide applications in many audio-related tasks [1], [5], [58]. Inspired by its successes in previous tasks, the encoder adopted in this work is the backbone of a ResNet18 [63]. It should be noted that we do not intentionally design an encoder with specialized structure and parameters. Instead, a common encoder is used here, since our method can be applicable and effective for various types of encoders.

When training the encoder, we only have the samples of base classes. The prototype of a base class is generated by calculating mean vector of embeddings of training samples. Owing to the sufficient samples per base class, the mean vector of embeddings can characterize the properties of each base class well. Augmented base-classes do not have samples, and thus their prototypes cannot be obtained in the same way as we generate prototypes of base classes. Their prototypes are generated by an algorithm which is described below.

The embedding $\boldsymbol{e}$ of a sample $\boldsymbol{x}$ of base classes is learned by the encoder $\boldsymbol{\theta}$, namely $\boldsymbol{e} = f_{\boldsymbol{\theta}}(\boldsymbol{x})$. The prototypes of the $h$th base class and augmented base-class are denoted as $\boldsymbol{p}_h^{bc}$ and $\boldsymbol{p}_h^{ab}$, respectively, where $1 \le h \le H$ and $H$ is the number of base classes (or augmented base-classes). The prototypes of $H$ base classes and $H$ augmented base-classes are denoted as $\boldsymbol{P}^{bc} = \{\boldsymbol{p}_h^{bc}\}_{1\le h\le H}$ and $\boldsymbol{P}^{ab} = \{\boldsymbol{p}_h^{ab}\}_{1\le h\le H}$, respectively. $\boldsymbol{p}_h^{bc}$ is calculated by

$$\boldsymbol{p}_h^{bc} = \frac{1}{Z_h}\sum_{j=1}^{Z_h} \boldsymbol{e}_{h,j} \, , \quad (3)$$

where $Z_h$ and $\boldsymbol{e}_{h,j}$ denote the number of samples from the $h$th base class and the embedding of the $j$th sample from the $h$th base class, respectively.

The distance between $\boldsymbol{e}$ and $\boldsymbol{p}_h^{ab}$ is defined by

$$d(\boldsymbol{e}, \boldsymbol{p}_h^{ab}) = \left\| \boldsymbol{e} - \boldsymbol{p}_h^{ab} \right\|_2^2 . \quad (4)$$

The distance $d(\boldsymbol{e}, \boldsymbol{p}_h^{ab})$ is used to determine which base class the sample $\boldsymbol{x}$ belongs to. The probability of sample $\boldsymbol{x}$ belongs to base class $h$ is proportional to the distance $d(\boldsymbol{e}, \boldsymbol{p}_h^{ab})$, namely $P(y = h|\boldsymbol{x}) \propto d(\boldsymbol{e}, \boldsymbol{p}_h^{ab})$. The larger the distance $d(\boldsymbol{e}, \boldsymbol{p}_h^{ab})$ is, the larger the probability that sample $\boldsymbol{x}$ is recognized as base class $h$. Given the encoder $\boldsymbol{\theta}$ and prototypes $\boldsymbol{P}^{ab}$, the probability that sample $\boldsymbol{x}$ is recognized as base class $h$ is defined by

$$P(y = h|\boldsymbol{x}, \boldsymbol{\theta}, \boldsymbol{P}^{ab}) = \frac{\exp\left(d(\boldsymbol{e}, \boldsymbol{p}_h^{cb})\right)}{\sum_{l=1}^{H} \exp\left(d(\boldsymbol{e}, \boldsymbol{p}_l^{cb})\right)} . \quad (5)$$

A loss of empirical classification risk is defined as the negative log-probability of the true base class $h$, namely

$$\mathcal{L}^c(\boldsymbol{x};\, \boldsymbol{\theta}, \boldsymbol{P}^{ab}) = -logP(y = h|\boldsymbol{x}, \boldsymbol{\theta}, \boldsymbol{P}^{ab}) . \quad (6)$$

To ensure that the prototype of each augmented base-class has a significant interval away from the prototype of its corresponding base class in the prototype space, a loss of complementary space risk is defined by

$$\mathcal{L}^s(\boldsymbol{x};\, \boldsymbol{\theta},\, \boldsymbol{p}_h^{ab}, M_h) = \left\| d(\boldsymbol{e}, \boldsymbol{p}_h^{ab}) - M_h \right\|_2^2 , \quad (7)$$

where $M_h$ is a learnable margin for the prototype of the $h$th augmented base-class, and $1 \le h \le H$. Minimizing the loss in Eq. (7) will result in an interval between the prototype of every augmented base-class and the prototype of its corresponding base class without overlapping.

Finally, a joint loss is defined as the combination of the losses of both empirical classification risk and complementary space risk, namely

$$\mathcal{L}(\boldsymbol{x};\, \boldsymbol{\theta}, \boldsymbol{P}^{ab}, \boldsymbol{M}) = \mathcal{L}^c(\boldsymbol{x};\, \boldsymbol{\theta}, \boldsymbol{P}^{ab}) + \alpha\times\mathcal{L}^s(\boldsymbol{x};\, \boldsymbol{\theta}, \boldsymbol{P}^{ab}, \boldsymbol{M}) , \quad (8)$$

where $\alpha$ is a weighting coefficient and $\boldsymbol{M} = \{M_h\}_{1\le h\le H}$ is a vector of learnable margins. $\boldsymbol{P}^{ab}$ and $\boldsymbol{M}$ are the two learnable parameters of the proposed algorithm for generating the prototypes of augmented base-classes. The details of the proposed algorithm are presented in Table III.

TABLE III
THE PROPOSED ALGORITHM FOR GENERATING THE PROTOTYPES OF AUGMENTED BASE-CLASSES.

**Input**: Training set of base classes $\{\boldsymbol{x}_{h,j}\}$, $1 \le h \le H$, $1 \le j \le Z_h$. Initialized encoder $\boldsymbol{\theta}$. Randomly initialized prototypes $\boldsymbol{P}^{ab} = \{\boldsymbol{p}_h^{ab}\}_{1\le h\le H}$ and margin vector $\boldsymbol{M}=\{M_h\}_{1\le h\le H}$. Weighting coefficient $\alpha$, learning rate $r$ and number of iterations $t = 0$.
**While** not converge **do**:
  Calculate the joint loss $\mathcal{L}_t = \mathcal{L}_t^c + \alpha\times\mathcal{L}_t^s$ , namely the loss of Eq. (8).
  Calculate backpropagation error $\frac{\partial\mathcal{L}_t}{\partial\boldsymbol{x}_{h,j}} = \frac{\partial\mathcal{L}_t^c}{\partial\boldsymbol{x}_{h,j}} + \alpha\times\frac{\partial\mathcal{L}_t^s}{\partial\boldsymbol{x}_{h,j}}$ .
  Update prototypes $\boldsymbol{p}_{h,t+1}^{ab} = \boldsymbol{p}_{h,t}^{ab} - r_t\times(\frac{\partial\mathcal{L}_t^c}{\partial\boldsymbol{p}_{h,t}^{ab}} + \alpha\times\frac{\partial\mathcal{L}_t^s}{\partial\boldsymbol{p}_{h,t}^{ab}})$ .
  Update margin vector $M_{h,t+1} = M_{h,t} - \alpha\times r_t\times\frac{\partial\mathcal{L}_t^s}{\partial M_{h,t}}$ .
  Update encoder $\boldsymbol{\theta}_{t+1} = \boldsymbol{\theta}_t - r_t\times(\frac{\partial\mathcal{L}_t^c}{\partial\boldsymbol{\theta}_t} + \alpha\times\frac{\partial\mathcal{L}_t^s}{\partial\boldsymbol{\theta}_t})$ .
  $t = t + 1$.
**End while**
**Output**: The prototypes $\boldsymbol{P}^{ab}$, margin vector $\boldsymbol{M}$ and encoder $\boldsymbol{\theta}$.

### *D. Prototype Generator of Few-shot Class*

The information of query and support embeddings is useful for enhancing the ability of the prototypes of few-shot classes to recognize the class to which each query embedding belongs. To generate the prototypes of few-shot classes by aggregating the information of query and support embeddings, we design a PGFC, including three modules: a Support-query (support & query) Embeddings Fusion Module (SEFM), a Support-embeddings Conversion Module (SCM), and a Salient-information Aggregation Module (SAM), as shown in Fig. 5.

First, the SEFM is designed as an attention module of the Transformer [64] in structure, and can capture the relationship among support embeddings $\boldsymbol{e}^s$ and query embeddings $\boldsymbol{e}^q$. The information of $\boldsymbol{e}^s$ and $\boldsymbol{e}^q$ can be fused and integrated into the transformed query embeddings $\hat{\boldsymbol{e}}^q$. Concretely, $\boldsymbol{e}^s$ is converted by two separate Conv (1×1 convolution, like a linear operation) to obtain vectors $\boldsymbol{e}_1$ and $\boldsymbol{e}_2$, and $\boldsymbol{e}^q$ is converted by a Conv to obtain vector $\boldsymbol{e}_3$. After conducting vector multiplication and Softmax on $\boldsymbol{e}_1$ and $\boldsymbol{e}_2$, $\boldsymbol{e}_{1,2}$ and $\boldsymbol{e}'_{1,2}$ are obtained. The element-wise product of $\boldsymbol{e}'_{1,2}$ and $\boldsymbol{e}_3$, namely $\boldsymbol{e}'_{1,2,3}$, is fed to a Conv to obtain $\hat{\boldsymbol{e}}^q$. The transformed query embeddings $\hat{\boldsymbol{e}}^q$ is defined by

$$\hat{\boldsymbol{e}}^q = Conv\left(\text{Softmax}\left(Conv(\boldsymbol{e}^s)(Conv(\boldsymbol{e}^q))^{\text{T}}\right)Conv(\boldsymbol{e}^q)\right), \quad (9)$$

where T and $Conv(\cdot)$ denote transpose operation of a matrix and 1×1 convolution, respectively.

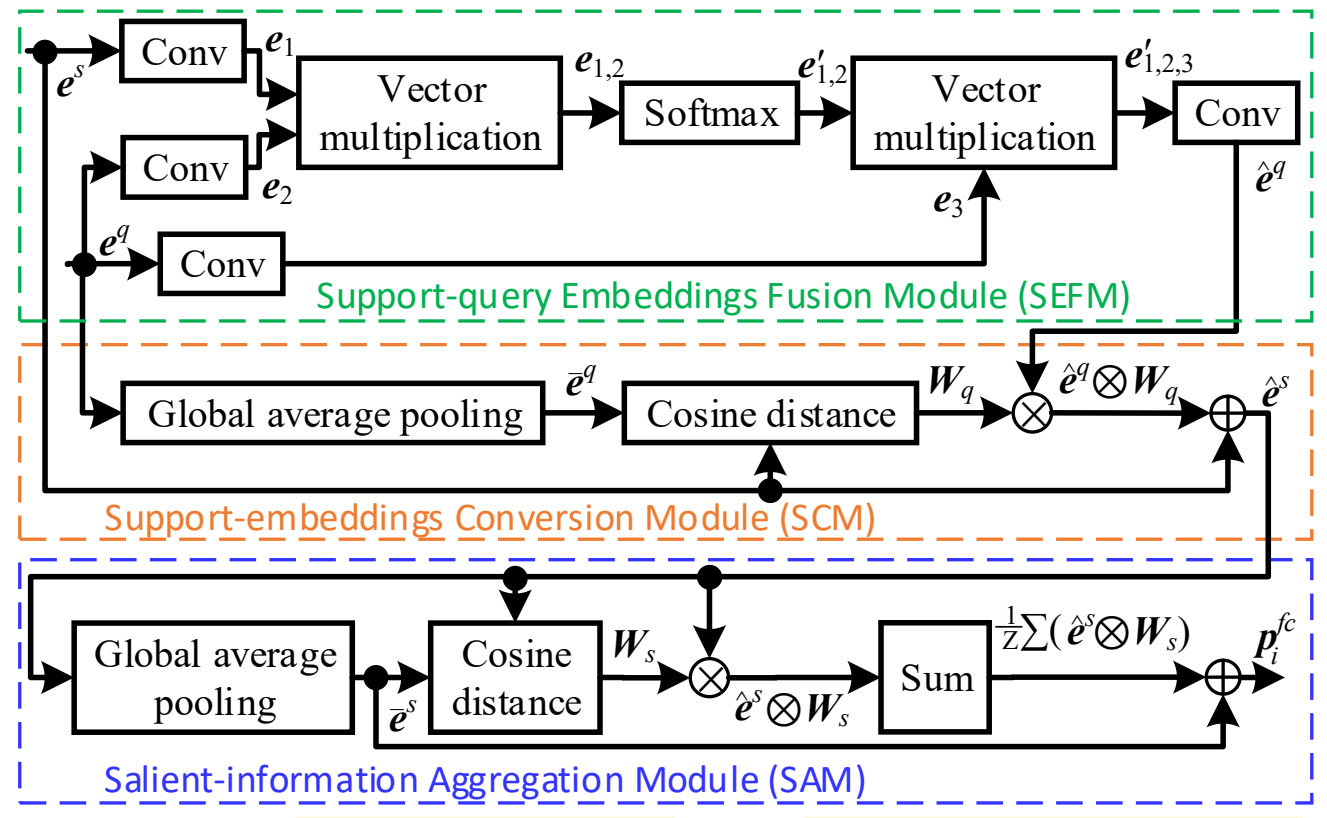

Fig. 5. Structure of the prototype generator of few-shot class which consists of a SEFM, a SCM and a SAM. The SEFM's input and output are support embeddings $\boldsymbol{e}^s$ and transformed query embeddings $\hat{\boldsymbol{e}}^q$, respectively. The SCM's input includes $\boldsymbol{e}^s$, $\hat{\boldsymbol{e}}^q$ and query embeddings $\boldsymbol{e}^q$, while its output is transformed support embedding $\hat{\boldsymbol{e}}^s$. The input and output of the SAM are $\hat{\boldsymbol{e}}^s$ and the $i$th few-shot class's prototype $\boldsymbol{p}_i^{fc}$, respectively.

Then, the SCM is designed for generating the transformed support embeddings $\hat{\boldsymbol{e}}^s$ by incorporating information from $\hat{\boldsymbol{e}}^q$. It predicts the parts of $\boldsymbol{e}^s$ which are greatly consistent with $\boldsymbol{e}^q$ as weighting coefficients vector $\boldsymbol{W}_q$. After performing the Hadamard product $\hat{\boldsymbol{e}}^q \otimes \boldsymbol{W}_q$ and the element-wise summation $(\hat{\boldsymbol{e}}^q \otimes \boldsymbol{W}_q) \oplus \boldsymbol{e}^s$, the information from $\boldsymbol{e}^s$ and $\hat{\boldsymbol{e}}^q$ flows into $\hat{\boldsymbol{e}}^s$. In short, the transformed support embeddings $\hat{\boldsymbol{e}}^s$ is defined by

$$\hat{\boldsymbol{e}}^s = \boldsymbol{e}^s \oplus \left(\hat{\boldsymbol{e}}^q \otimes Cos(GAP(\boldsymbol{e}^q),\ \boldsymbol{e}^s)\right), \quad (10)$$

where $Cos(\cdot,\cdot)$ and $GAP(\cdot)$ denote cosine distance and global average pooling, respectively; $\oplus$ and $\otimes$ represent element-wise summation and Hadamard product, respectively.

Finally, the SAM is designed to preserve salient information of each transformed support embedding $\hat{\boldsymbol{e}}^s$ by building local-to-global correlation. The discriminative and salient information in each sample can be preserved and aggregated into the $i$th few-shot class's prototype $\boldsymbol{p}_i^{fc}$. Specifically, $\hat{\boldsymbol{e}}^s$ is transformed by an operation of global average pooling to obtain $\bar{\boldsymbol{e}}^s$. After computing a cosine distance between $\hat{\boldsymbol{e}}^s$ and $\bar{\boldsymbol{e}}^s$, the weighting coefficients vector $\boldsymbol{W}_s$ is obtained. The salient information of $\hat{\boldsymbol{e}}^s$ is aggregated into the prototype $\boldsymbol{p}_i^{fc}$ by sequentially executing the operations of Hadamard product, sum and element-wise summation. In short, $\boldsymbol{p}_i^{fc}$ is obtained by

$$\boldsymbol{p}_i^{fc} = GAP(\hat{\boldsymbol{e}}^s) \oplus \left(\frac{1}{Z}\sum\left(\hat{\boldsymbol{e}}^s \otimes Cos(GAP(\hat{\boldsymbol{e}}^s),\ \hat{\boldsymbol{e}}^s)\right)\right), \quad (11)$$

where $Z$ and $\Sigma$ denote the number of vectors for normalization and the sum operation of vectors, respectively.

## E. *Prototype Generator of Open-set Class*

Support samples of open-set classes are unavailable in practice. The open-set class's prototype $\boldsymbol{p}^{oc}$ cannot be generated by the support samples of open-set class. We generate $\boldsymbol{p}^{oc}$ by a PGOC which is composed of two attention modules and one post-processing module, as shown in Fig. 6.

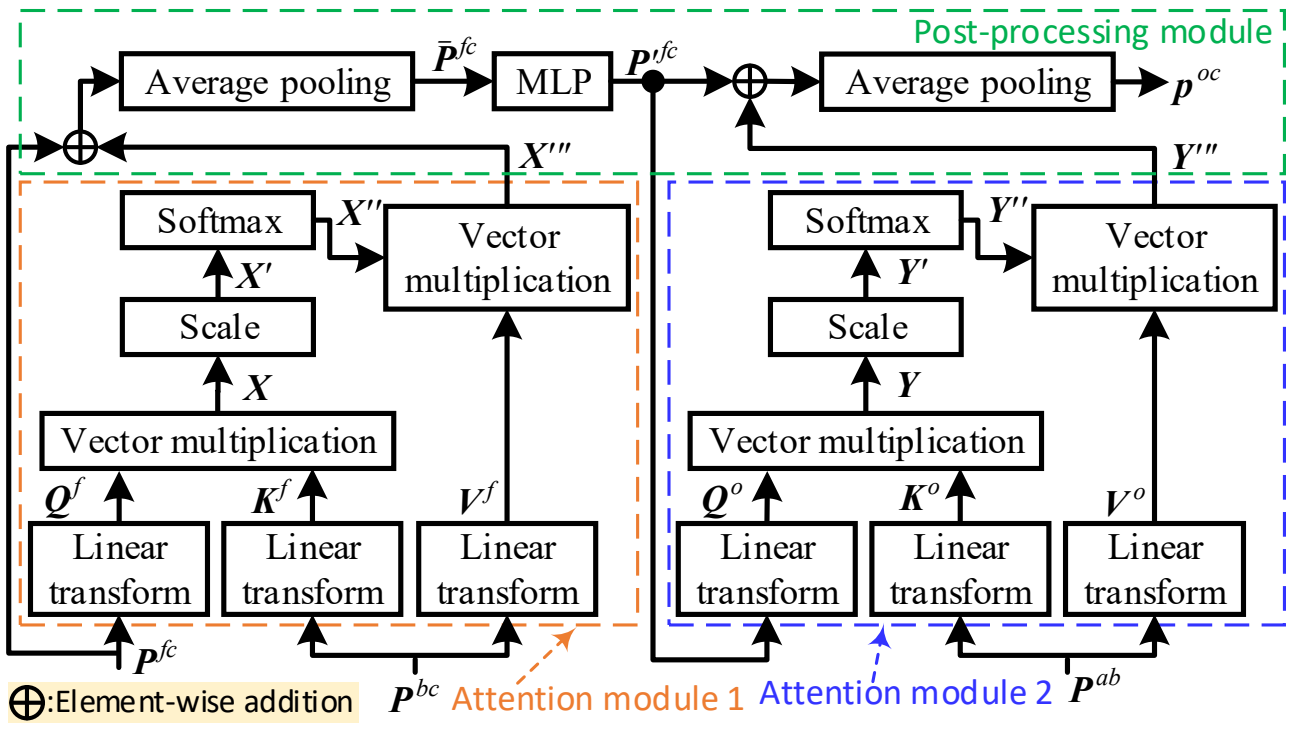

Fig. 6. Structure of the proposed prototype generator of open-set class.

The attention module of Transformer [64] and the Multi-Layer Perceptron (MLP) [36] are proved to be effective in exploiting the relations of input vectors and in generating negative prototype of input prototype, respectively. We use the attention modules of Transformer and MLP as main parts of the PGOC whose structure design is based on the following two considerations. First, the few-shot class's prototypes $\boldsymbol{P}^{fc}$ and base class's prototypes $\boldsymbol{P}^{bc}$ have the same property as they are generated by embeddings, and are fed to the same attention module to obtain attentive prototype $\boldsymbol{X}'''$. Second, $\boldsymbol{P}'^{fc}$ is converted from mean prototype $\bar{\boldsymbol{P}}^{fc}$ by the MLP, and is called negative prototype of $\boldsymbol{P}^{fc}$. $\boldsymbol{P}'^{fc}$ and $\boldsymbol{P}^{ab}$ have the same property as they are negative prototypes or augmented prototypes (complementary to prototypes of base classes), and thus are fed to the same attention module to generate attentive prototype $\boldsymbol{Y}'''$. The generation of $\boldsymbol{p}^{oc}$ is described below.

In the first step, we generate $\boldsymbol{P}'^{fc}$ with the attention module 1 and post-processing module. First, the operations of linear transform, vector multiplication, scale, Softmax and vector multiplication are conducted on $\boldsymbol{P}^{fc}$ and $\boldsymbol{P}^{bc}$ for sequentially obtaining the vectors of $\boldsymbol{X} = \boldsymbol{Q}^f(\boldsymbol{K}^f)^{\text{T}}$, $\boldsymbol{X}' = \boldsymbol{X}/\sqrt{D}$, $\boldsymbol{X}'' = \text{Softmax}(\boldsymbol{X}')$ and $\boldsymbol{X}''' = \boldsymbol{X}''\boldsymbol{V}^f$. T and $D$ denote the operation of matrix transpose and dimension of the input prototype. Then, mean prototype $\bar{\boldsymbol{P}}^{fc}$ is obtained by performing average pooling on the element-wise addition of both $\boldsymbol{P}^{fc}$ and $\boldsymbol{X}'''$,

$$\bar{\boldsymbol{P}}^{fc} = AP\left(\boldsymbol{P}^{fc} \oplus \left(\text{Softmax}\left(\frac{1}{\sqrt{D}}\boldsymbol{Q}^f(\boldsymbol{K}^f)^{\text{T}}\right)\right)\boldsymbol{V}^f\right), \quad (12)$$

where $AP(\cdot)$ and $\oplus$ denote average pooling and element-wise addition, respectively. Finally, $\boldsymbol{P}'^{fc}$ is obtained by converting $\bar{\boldsymbol{P}}^{fc}$ using the MLP.

In the second step, $\boldsymbol{p}^{oc}$ is generated by the attention module 2 and post-processing module. Its generation is the same as that of $\bar{\boldsymbol{P}}^{fc}$ except for different inputs, and is denoted by

$$\boldsymbol{p}^{oc} = AP\left(\boldsymbol{P}'^{fc} \oplus \left(\text{Softmax}\left(\frac{1}{\sqrt{D}}\boldsymbol{Q}^o(\boldsymbol{K}^o)^{\text{T}}\right)\right)\boldsymbol{V}^o\right). \quad (13)$$

# IV. EXPERIMENTS

We first describe experimental datasets and setup. Then, we do comparison experiments in accuracy (including statistical significance test), complexity and generalizability to show the advantage of our method. Next, an ablation experiment is done to prove the effectiveness of main parts of our method. Finally,

TABLE IV
DATA DIVISIONS OF THE THREE EXPERIMENTAL DATASETS

| Datasets | Parameters | Training subset | | Testing subset | | |
|---|---|---|---|---|---|---|
| | | $\boldsymbol{D}_t^{bc}$ | $\boldsymbol{D}_t^{pn}$ | $\boldsymbol{D}_s^{fc}$ | $\boldsymbol{D}_q^{fc}$ | $\boldsymbol{D}_q^{oc}$ |
| FSC-89 | #Classes | 60 | 60 | 29 | 29 | 29 |
| | #Samples | 48000 | 48000 | 15000 | 15000 | 15000 |
| | Length (hours) | 13.33 | 13.33 | 4.17 | 4.17 | 4.17 |
| | #Samples/Class | 800 | 800 | 500 | 500 | 500 |
| NSynth-100 | #Classes | 60 | 60 | 40 | 40 | 40 |
| | #Samples | 12000 | 12000 | 4500 | 4500 | 4500 |
| | Length (hours) | 12.23 | 12.23 | 5.00 | 5.00 | 5.00 |
| | #Samples/Class | 200 | 200 | 100 | 100 | 100 |
| LS-100 | #Classes | 60 | 60 | 40 | 40 | 40 |
| | #Samples | 30000 | 30000 | 20000 | 20000 | 20000 |
| | Length (hours) | 16.66 | 16.66 | 11.11 | 11.11 | 11.11 |
| | #Samples/Class | 500 | 500 | 500 | 500 | 500 |

#Samples/Class: number of samples for each class.

we do three extended experiments to show the impacts of the *N*-way *K*-shot values, proportion of open-set classes' samples and classifier on the performance of our method.

### *A. Experimental Datasets*

We adopt FSC-89, NSynth-100 and LS-100 as experimental datasets to verify the effectiveness all methods. These three datasets have been widely applied in previous work. To reimplement the results of this work, these three datasets can be publicly downloaded from the websites[1]. The FSC-89 has 89 types of sounds emitted by humans, animals, nature and objects. The NSynth-100 has types of 100 sounds emitted by musical instruments. The LS-100 is a speech corpus, consisting of utterances of 100 speakers.

Each dataset is divided into two subsets: training and testing subsets. The training subset is divided into $\boldsymbol{D}_t^{bc}$ (training set of base classes) and $\boldsymbol{D}_t^{pn}$ (training set of pseudo novel-classes). The testing subset is divided into $\boldsymbol{D}_s^{fc}$ (support set of few-shot classes) and $\boldsymbol{D}_q = \boldsymbol{D}_q^{fc} \cup \boldsymbol{D}_q^{oc}$ (query set of novel classes). The $\boldsymbol{D}_t^{bc}$ is used to pre-train the model in supervised manner. The $\boldsymbol{D}_t^{pn}$ is used to meta-train the model in episodic way. The $\boldsymbol{D}_s^{fc}$ and $\boldsymbol{D}_q$ are used to meta-train and test the model, respectively. Data divisions of the three datasets are given in Table IV.

### *B. Experimental Setup*

All experiments are conducted on a high-performance computer which has two Intel CPUs of Xeon-8124M with 3.50 GHz, a RAM with 128 GB, and three NVIDIA 3090 GPUs. Accuracy (Acc) and Area Under the Receiver Operating Characteristic curve (AUROC) are used as main performance metrics. Acc is defined as the number of correctly recognized samples divided by the total number of samples involved in classification, which is used to measure the classification performance of various methods on all seen classes. Acc is computed only on seen-class samples by taking the argmax of the model's output over the seen classes. It does not require a threshold of out-of-distribution detection, as it evaluates standard classification performance within the seen label space. AUROC is calculated as the area under the receiver operating characteristic curve which shows the trade-off between true and false positive rates across various decision thresholds. It can be used to measure the performance of different methods in distinguishing between seen and unseen classes. It is threshold-independent and reflects the separability of seen versus unseen data. The metrics of Multiply-ACcumulate operations (MACs) and Average Inference Time (AIT) for each query sample are used to measure the computational complexity of different methods, while Number of Parameters (NP) during inference is used to measure the memory footprint of various methods. The lower the MACs, AIT and NP, the lower complexity of the methods. Main parameters of our method are listed in Table V.

[1] https://www.modelscope.cn/datasets/pp199124903/LS-100/summary
https://www.modelscope.cn/datasets/pp199124903/FSC-89/summary
https://www.modelscope.cn/datasets/pp199124903/NSynth-100/summary

TABLE V
SETTINGS OF MAIN PARAMETERS OF OUR METHOD

| Parameters | Values | Parameters | Values |
|---|---|---|---|
| Frame length/shift | 25/10 ms | Learning rate | 0.001 |
| Dimension of embedding, $D$ | 512 | Weight decay | 0.0005 |
| Dimension of log Mel-spectrum | 80 | Training epochs | 50 |
| No. of ways per episode, $N$ | 1 to 15 | Training episodes | 300 |
| No. of support sample, $K$ | 1 to 15 | No. of test runs | 600 |
| No. of query sample for each FC | 15 | Coefficient $q_\varsigma$ | 3.102 |
| No. of query sample for each OC | 15 | Weighting coef., $\alpha$ | 1.0 |

FC: Few-shot Class; OC: Open-set Class.

TABLE VI
TECHNICAL FEATURES OF DIFFERENT METHODS

| Methods | Main features |
|---|---|
| FEAT | Few-shot embedding adaptation with Transformer |
| L$^3$-Net | Self-supervised learning without requiring labeled data |
| D-ProtoNet | Metric learning, episode-known dummy prototypes |
| OpenFEAT | Embeddings adaption using Transformer-based set-to-set function |
| TANE | Augments closed-set classifier with negative prototypes |
| GEL | Energy-based hybrid model; fusing global and local information |
| OPP | An overall positive prototype is used to represent positive samples |
| MET | Meta evidential Transformer; evidential open-set loss |
| Ours | Attention information-fused prototypes |

Main parameters of previous methods are set according to the suggestions of their original papers and optimally adjusted on the experimental datasets. Consistent with the majority of previous methods, the setting of *N*-way *K*-shot is 5-way 1-shot and 5-way 5-shot in the subsequent experiments of this paper, unless otherwise specified.

### *C. Comparison in Accuracy and AUROC*

In this subsection, we compare our method to prior methods in Acc and AUROC. The abbreviations of eight prior methods are FEAT [37], L$^3$-Net [23], D-ProtoNet [21], OpenFEAT [22], TANE [36], GEL [38], OPP [39], and MET [40].

Table VI briefly gives technical features of various methods. Previous methods are reimplemented with the code released by the papers or written by us based on the descriptions of the papers. Experimental results obtained by various methods on the LS-100, NSynth-100 and FSC-89 are listed in Table VII. We can draw two conclusions from the results in Table VII.

First, the Acc and AUROC scores obtained by all methods on the FSC-89 are consistently lower than those on both the LS-100 and NSynth-100. The reasons for this result are that the samples in the FSC-89 contain stronger noises and have poorer intra-class consistency.

Second, our method exceeds other methods in Acc and AUROC on all datasets under the settings of 5-way 1-shot and 5-way 5-shot. Specifically, it is known that our method obtains Acc score of 85.98% (or 95.25%) on the LS-100, Acc score of 88.05% (or 94.30%) on the NSynth-100, and Acc score of 63.62% (or 80.83%) on the FSC-89, when the setting of $N$-way $K$-shot is 5-way 1-shot (or 5-way 5-shot). These Acc scores obtained by our method are higher than the counterparts achieved by other methods. In addition, our method obtains AUROC score of 90.55% (or 92.38%) on the LS-100, AUROC score of 85.67% (or 91.26%) on the NSynth-100, and AUROC score of 71.24% (or 82.06%) on the FSC-89, when the setting of $N$-way $K$-shot is 5-way 1-shot (or 5-way 5-shot). These AUROC scores achieved by our method are also higher than the counterparts obtained by other methods.

TABLE VII
RESULTS OBTAINED BY DIFFERENT METHODS ON THREE DATASETS (IN %)

| Methods | LS-100 | | | | NSynth-100 | | | | FSC-89 | | | |
|---|---|---|---|---|---|---|---|---|---|---|---|---|
| | 5-way 1-shot | | 5-way 5-shot | | 5-way 1-shot | | 5-way 5-shot | | 5-way 1-shot | | 5-way 5-shot | |
| | Acc | AUROC | Acc | AUROC | Acc | AUROC | Acc | AUROC | Acc | AUROC | Acc | AUROC |
| FEAT [37] | 85.36 | 82.18 | 94.39 | 86.86 | 85.62 | 71.84 | 89.92 | 81.41 | 62.85 | 68.77 | 80.50 | 80.25 |
| $L^3$-Net [23] | 44.12 | 55.34 | 89.17 | 74.93 | 64.99 | 67.80 | 92.19 | 85.58 | 42.75 | 55.16 | 73.33 | 59.47 |
| D-ProtoNet [21] | 84.44 | 83.54 | 90.70 | 77.31 | 86.40 | 83.57 | 94.19 | 84.34 | 52.83 | 63.32 | 72.50 | 63.42 |
| OpenFEAT [22] | 80.50 | 85.86 | 89.21 | 91.73 | 62.17 | 74.03 | 75.86 | 81.88 | 30.08 | 53.70 | 50.72 | 63.96 |
| TANE [36] | 85.74 | 78.89 | 94.95 | 84.14 | 84.87 | 77.21 | 92.49 | 80.65 | 63.31 | 70.72 | 80.80 | 79.38 |
| GEL [38] | 84.47 | 84.95 | 93.92 | 92.00 | 67.93 | 75.50 | 82.44 | 84.48 | 32.21 | 56.59 | 58.65 | 67.73 |
| OPP [39] | 84.11 | 79.79 | 94.80 | 81.95 | 83.05 | 76.03 | 90.57 | 77.02 | 62.36 | 69.23 | 80.46 | 74.44 |
| MET [40] | 84.46 | 83.82 | 89.44 | 87.52 | 85.42 | 83.81 | 92.56 | 88.76 | 60.85 | 69.79 | 75.80 | 79.08 |
| Ours | **85.98** | **90.55** | **95.25** | **92.38** | **88.05** | **85.67** | **94.30** | **91.26** | **63.62** | **71.24** | **80.83** | **82.06** |

The above gains of both Acc and AUROC scores obtained by our method are mainly due to the usages of base classes' prototypes, PGFC and PGOC for the prototypes' generation of novel classes. First, the prototypes of base classes possess strong representational ability since they are generated by abundant samples of base classes. Second, the PGFC generates prototypes of few-shot classes through fusing the class-discriminative information of support and query embeddings and assigning larger weighting coefficient to representative part of the support embeddings. Third, the PGOC generates prototypes of open-set classes via integrating the prototypes of few-shot classes and base classes. Based on the above three measures, the generated prototypes of novel classes possess strong distinguishing and generalization abilities. Hence, our method effectively distinguishes between few-shot and open-set classes (higher AUROC scores), and accurately recognize each few-shot class (higher Acc scores).

## D. *Statistical Significance Test*

From the results in the last subsection, we know that our method has advantage over prior methods in Acc and AUROC. In this subsection, we discuss whether the above advantage of our method is of statistical significance. The distributions of samples in the three experimental datasets are unknown, and the samples are not assumed to follow a specific distribution. We utilize a non-parametric method for statistical significance test, because the distribution of samples is not required to be known in advance for the non-parametric method. To realize the statistical significance test, we use the Friedman test and the Nemenyi test [65] as the null-hypothesis test and the post-hoc test, respectively. The combination of the Friedman test and the Nemenyi test has been popularly adopted when the distribution of samples is unknown.

We assess all methods on various data subsets. The methods that have the first, second and third highest AUROC or Acc scores on a data subset are ranked as 1, 2 and 3, respectively, and so on. Let $r_b^c$ denote the ranking achieved by the $c$th method on the $b$th data subset. The average ranking $R_c$ yielded

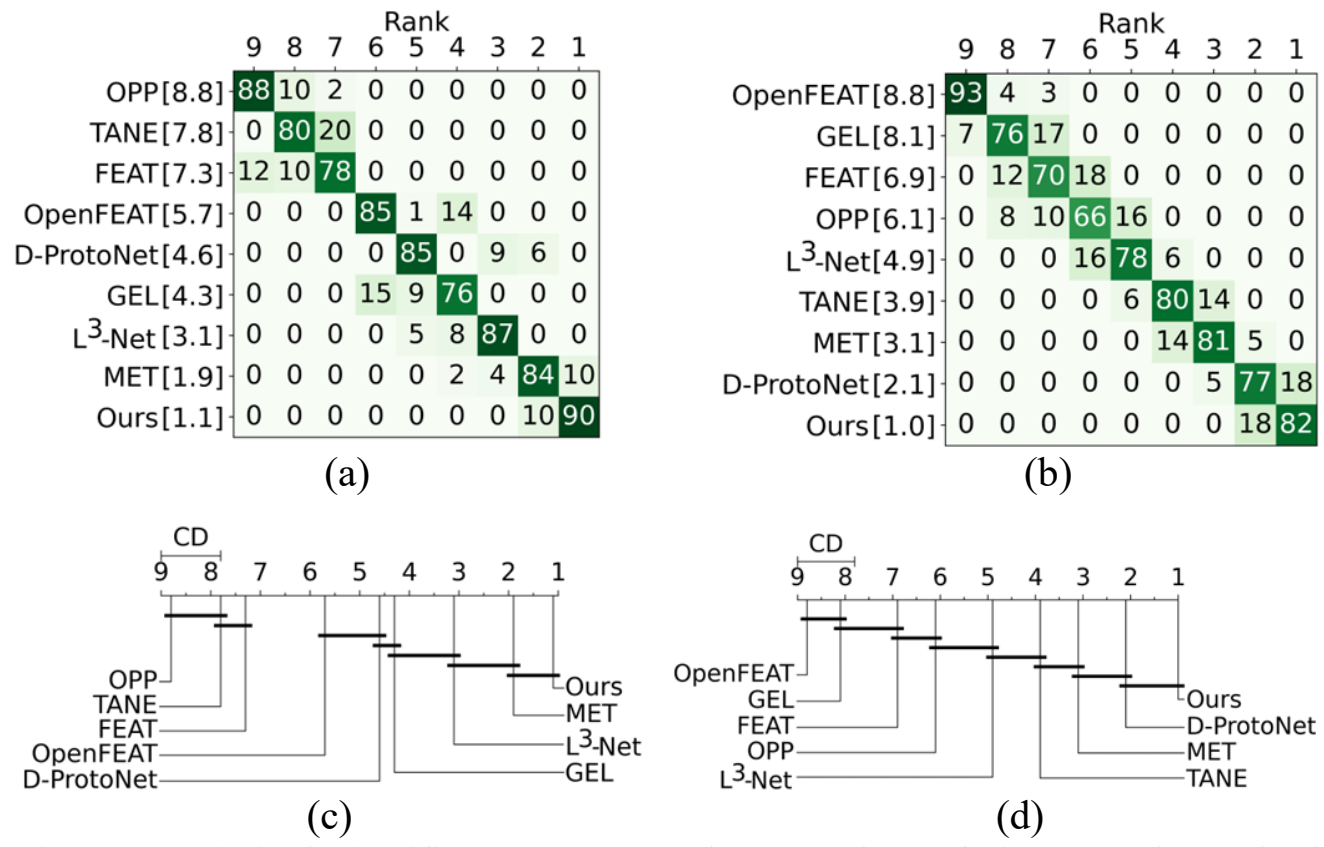

Fig. 7. Statistical significance test results. Number of times each method obtains each ranking for computing statistical significance differences in: (a) AUROC and (b) Acc. Statistical significance differences for all methods in terms of: (c) AUROC and (d) Acc.

by the $c$th method on $B$ data subsets is defined by

$$R_c = \frac{1}{B}\sum_{b=1}^{B} r_b^c \ . \tag{14}$$

where $1 \leq c \leq C$, $1 \leq b \leq B$; $C$ and $B$ represent the number of methods and the number of data subsets, respectively. In the null-hypothesis, all methods have the same $R_c$ value and thus there is no statistically significant discrepancy between them. If the above null-hypothesis is refused by the results of the Friedman test, the statistically significant discrepancy between all methods is decided by the Neminyi test. If the discrepancy of $R_c$ values between any two methods exceeds a critical difference (CD), then the statistical significance discrepancy in AUROC or Acc occurs between these two methods [65]. The definition of CD is represented by

$$CD = q_\varsigma\sqrt{\frac{C(C+1)}{6B}}\ , \tag{15}$$

where $q_\varsigma$ denotes a coefficient which is determined based on the Studentized range statistic multiplied by $1/\sqrt{2}$ [65].

We randomly choose samples from the LS-100, NSynth-100, or FSC-89 to construct 100 data subsets for evaluating our method and other eight methods. Hence, $C$ = 9 and $B$ = 100. We rank all methods in accordance to their AUROC or Acc scores that are achieved on each data subset. Hence, there are 100 rankings for each method. Fig. 7 shows the statistical significance test results for all methods.

Fig. 7 (a) and (b) illustrates number of times each method obtains each ranking for computing statistical significance differences in terms of (a) AUROC and (b) Acc, respectively. All methods are arranged according to their average rankings (decimals in the square brackets on the left side of Fig. 7 (a) and (b)). If the average ranking of a method is smaller, then the performance of this method is better. For instance, our method obtains the best performance among all methods since it has the smallest average ranking of 1.1 in AUROC (shown in Fig. 7 (a)) and 1.0 in Acc (shown in Fig. 7 (b)). The four best performing methods in AUROC are ours, MET, $L^3$-Net and GEL, while the four best performing methods in Acc are ours, D-ProtoNet, MET and TANE.

TABLE VIII
THE VALUES OF MACS, AIT AND NP OF DIFFERENT METHODS

| Methods | MACs (million) | AIT (second) | NP (million) |
|---|---|---|---|
| FEAT [37] | 7241 | 4.41 | 12.38 |
| $L^3$-Net [23] | **1973** | **3.24** | **11.33** |
| D-ProtoNet [21] | **1973** | 3.40 | 11.37 |
| OpenFEAT [22] | 28213 | 6.63 | 11.34 |
| TANE [36] | 30102 | 7.14 | 15.86 |
| GEL [38] | 28268 | 7.54 | 11.59 |
| OPP [39] | 30103 | 7.66 | 24.49 |
| MET [40] | 7242 | 4.55 | 12.38 |
| Ours | 2259 | 3.37 | 15.02 |

Fig. 7 (c) and (d) show the statistical significance differences of all methods in AUROC and Acc, respectively. The confidence level ς is typically set to 0.05. In the abscissa axis, the digits from 1 to 9 denote average rankings of all methods. If the methods are not crossed by a black-thick crossbar, then the performance difference between them is of statistical significance; otherwise, the performance difference between them is not statistically significant. It is known from Fig. 7 (c) and (d) that our method has statistically significant advantage over previous methods except the MET method in AUROC, and except the D-ProtoNet method in Acc.

### E. Comparison in Complexity

Both MACs and AIT are adopted to measure computational complexity, and NP is used to measure the memory overhead of different methods. The values of MACs, AIT and NP of different methods are given in Table VIII.

Based on the MACs values of all methods, we can draw the following two conclusions. First, the MACs value of our method is 2259 million which is smaller than the counterparts of other methods except the $L^3$-Net and D-ProtoNet methods. Second, our method and the $L^3$-Net and D-ProtoNet methods have similar and smaller MAC values, but the MACs values of other six methods are larger. The main reasons are below. The three methods have only one branch in structure or don't have modules with heavy computational load. In converse, the other six methods have multiple branches or have computationally complex modules. For example, the GEL method contains two branches (class-wise and pix-wise branches) for similarity computation. The TANE method has a conjugate class training module. The OPP method adopts five modules of Transformer to generate overall positive and other prototypes. The above modules used in these methods are computationally complex.

Based on the AIT values of various methods, we can obtain the following two observations. First, the AIT value of our method is 3.37 seconds which is smaller than the counterparts of previous methods except the $L^3$-Net method. Second, the AIT values of various methods are basically proportional to their MACs values. If the AIT value of a method is larger, its MACs value is basically larger; vice versa. For example, the OPP method has the largest MACs and AIT values, while the $L^3$-Net method has the smallest MACs and AIT values.

TABLE IX
ACC SCORES OBTAINED BY VARIOUS METHODS ON THE COMBINATIONS OF TWO DIFFERENT DATASETS (IN %)

| Methods | FS→NS | FS→LS | NS→FS | NS→LS | LS→FS | LS→NS |
|---|---|---|---|---|---|---|
| FEAT [37] | 39.38 | 41.94 | 40.75 | 42.76 | 35.78 | 55.13 |
| $L^3$-Net [23] | 72.62 | 69.42 | 56.07 | 58.50 | 43.57 | 63.06 |
| D-ProtoNet [21] | 70.13 | 71.19 | 49.28 | 51.87 | 44.44 | 65.76 |
| OpenFEAT [22] | 21.65 | 54.37 | 25.64 | 38.28 | 32.60 | 48.82 |
| TANE [36] | 61.43 | 61.78 | 43.14 | 33.72 | 42.24 | 64.41 |
| GEL [38] | 66.10 | 42.36 | 39.70 | 53.27 | 41.05 | 64.11 |
| OPP [39] | 67.80 | 58.71 | 42.01 | 42.59 | 43.39 | 65.65 |
| MET [40] | 41.26 | 43.51 | 45.85 | 35.83 | 41.06 | 53.11 |
| Ours | **74.20** | **72.18** | **56.98** | **61.28** | **48.28** | **67.23** |

NS: NSynth-100; FS: FSC-89; LS: LS-100.

TABLE X
AUROC SCORES OBTAINED BY VARIOUS METHODS ON THE COMBINATIONS OF TWO DIFFERENT DATASETS (IN %)

| Methods | FS→NS | FS→LS | NS→FS | NS→LS | LS→FS | LS→NS |
|---|---|---|---|---|---|---|
| FEAT [37] | 30.82 | 42.18 | 50.26 | 54.33 | 30.17 | 56.32 |
| $L^3$-Net [23] | 63.47 | 60.97 | 45.21 | 41.82 | 43.57 | 62.90 |
| D-ProtoNet [21] | 58.16 | 63.03 | 46.70 | 53.92 | 42.03 | 65.19 |
| OpenFEAT [22] | 23.86 | 56.12 | 22.14 | 35.56 | 31.69 | 62.88 |
| TANE [36] | 51.38 | 53.79 | 53.62 | 22.19 | 38.40 | 54.62 |
| GEL [38] | 65.27 | 44.63 | 33.66 | 56.27 | 41.04 | 63.63 |
| OPP [39] | 50.91 | 52.14 | 52.77 | 50.98 | 40.98 | 52.68 |
| MET [40] | 36.38 | 37.08 | 43.89 | 33.22 | 42.52 | 60.00 |
| Ours | **66.97** | **63.11** | **54.94** | **57.63** | **44.13** | **65.35** |

From the NP values of different methods, we can make the following two conclusions. First, the NP value of our method is 15.02 million which is larger than its counterparts of prior methods except the methods of TANE and OPP. Second, the NP values of various methods are not proportional to their MACs and AIT values except the $L^3$-Net and OPP methods. The reason for this result is that some methods have many modules for embeddings extraction and prototypes generation but the computational load of these modules is relatively low. Hence, these methods have larger NP values but their values of MACs and AIT are smaller. For example, our method mainly consists of an embedding extractor, a PGFC and a PGOC, but these modules are not computationally complex.

In short, ranking all methods from low to high complexity, our method ranks third, second and seventh in MACs, AIT and NP, respectively. Compared to most of the prior methods, our method does not have an advantage in NP, but it has an advantage in MACs and AIT.

### F. Comparison in Generalizability

In all the above experiments, training and testing samples are from the same dataset. To compare different methods under the condition of across datasets, the dataset for selecting training samples is different from the dataset for choosing testing samples. That is, the training dataset is different from the testing dataset in this subsection. Without loss of generality, the way-shot is set to 5-way 5-shot. Tables IX and X list the Acc and AUROC scores of various methods on the combinations of two different datasets, respectively.

The three datasets can be adopted for selecting training or testing samples, and thus we obtain six different combinations of datasets. As demonstrated in the first row of Tables IX and X, the abbreviations to the left and right of each arrow represent training dataset and testing dataset, respectively. For instance, in the case of NS→LS, the training and testing datasets are the NSynth-100 and LS-100, respectively. In the cases of LS→LS, NS→NS and FS→FS the training and

TABLE XI
RESULTS OBTAINED BY OUR METHOD ON THREE DATASETS WITH EIGHT COMBINATIONS OF PGFC, PGOC AND PAPN (IN %)

| Cases | PGFC | PGOC | PAPN | LS-100 | | | | NSynth-100 | | | | FSC-89 | | | |
|---|---|---|---|---|---|---|---|---|---|---|---|---|---|---|---|
| | | | | 5-way 1-shot | | 5-way 5-shot | | 5-way 1-shot | | 5-way 5-shot | | 5-way 1-shot | | 5-way 5-shot | |
| | | | | Acc | AUROC | Acc | AUROC | Acc | AUROC | Acc | AUROC | Acc | AUROC | Acc | AUROC |
| ① | × | × | × | 80.11 | 70.05 | 80.99 | 82.88 | 80.11 | 70.76 | 90.22 | 82.33 | 60.50 | 65.78 | 75.44 | 74.30 |
| ② | × | × | √ | 84.79 | 74.86 | 81.34 | 84.81 | 85.44 | 81.60 | 91.79 | 86.35 | 61.97 | 66.22 | 78.27 | 76.00 |
| ③ | × | √ | × | 83.50 | 78.90 | 85.56 | 85.50 | 82.23 | 83.22 | 91.12 | 88.32 | 61.33 | 66.59 | 76.40 | 75.66 |
| ④ | × | √ | √ | 84.82 | 80.69 | 86.33 | 85.91 | 85.67 | 84.31 | 92.81 | 90.50 | 62.46 | 67.98 | 78.78 | 80.01 |
| ⑤ | √ | × | × | 83.97 | 71.41 | 84.77 | 83.33 | 83.10 | 74.33 | 92.97 | 84.73 | 62.70 | 66.30 | 77.45 | 75.23 |
| ⑥ | √ | × | √ | 84.94 | 80.62 | 88.29 | 87.90 | 86.41 | 81.61 | 93.73 | 86.42 | 62.99 | 67.13 | 78.45 | 79.32 |
| ⑦ | √ | √ | × | 85.10 | 82.72 | 93.64 | 90.89 | 86.64 | 84.40 | 94.04 | 89.94 | 63.11 | 68.42 | 78.45 | 76.20 |
| ⑧ | √ | √ | √ | **85.98** | **90.55** | **95.25** | **92.38** | **88.05** | **85.67** | **94.30** | **91.26** | **63.62** | **69.22** | **78.93** | **82.06** |

×/√: the part isn't/is used. PGFC: Prototype Generator of Few-shot Class. PGOC: Prototype Generator of Open-set Class. PAPN: Prototypes of Augmented base-classes for generating Prototypes of Novel-classes. In cases ③ and ⑦, PGOC is used but PAPN isn't used (i.e., PGOC without PAPN), where all modules of PGOC in Fig. 6 are still in use and $\boldsymbol{P}^{ab}$ is replaced with $\boldsymbol{P}^{bc}$ for generating $\boldsymbol{p}^{oc}$ because both $\boldsymbol{P}^{ab}$ and $\boldsymbol{P}^{bc}$ are generated by samples of base classes.

testing samples are selected from the same dataset of LS-100, NSynth-100 and FSC-89, respectively. That is, the above three cases are not across datasets. As listed in Table VII, when the setting of *N*-way *K*-shot is 5-way 5-shot, our method obtains Acc (or AUROC) scores of 95.25% (or 92.38%), 94.30% (or 91.26%) and 80.83% (or 82.06%) in the cases of LS→LS, NS→NS and FS→FS, respectively. From the results in Tables VII, IX and X, the following two conclusions can be drawn.

First, the six Acc scores and the six AUROC scores obtained by our method are consistently higher than their counterparts achieved by other methods in Tables IX and X. From the results in Table VII, it is also known that our method obtains higher Acc and AUROC scores than other methods.

Second, the Acc and AUROC scores obtained by every method without crossing datasets are always higher than the counterparts obtained by this method in the cases of across datasets. For example, the Acc score of 80.83% obtained by our method in the case of FS→FS (without crossing datasets) is higher than 74.20% and 72.18% obtained by our method in the cases of FS→NS and FS→LS (with crossing datasets), respectively. Similarly, the AUROC score of 82.06% obtained by our method in the case of FS→FS is higher than 66.97% and 63.11% obtained by our method in the case of FS→NS and FS→LS, respectively. That is, when the training and testing samples are from different datasets, the Acc and AUROC scores obtained by all methods consistently decrease.

In conclusion, our method obtains higher Acc and AUROC scores than other methods under the condition of across datasets. Hence, it possesses stronger generalization ability than other methods across datasets. It should be noted that the factor of across datasets has a significant impact on the decline of Acc and AUROC scores for all methods.

### G. Ablation Experiments

In this subsection, we show contribution of each main part of our method. As shown in Fig. 4, novelties of our method are three-fold, including Prototypes of Augmented base-classes for generating Prototypes of Novel-classes (PAPN); a PGFC for obtaining prototypes of few-shot classes; and a PGOC for obtaining prototypes of open-set classes. To show contribution of the three parts to the performance gains, we implement our method with eight combinations of them. By adding one of the three parts, we can obtain its impact on the performance of our method. Table XI lists results obtained by our method on three datasets with eight combinations of PGFC, PGOC and PAPN.

Both Acc and AUROC scores obtained by our method with PGFC are higher than the counterparts obtained by our method without PGFC on three datasets under two way-shot settings. For instance, it can be seen from the results of cases ② and ⑥ that our method with PGFC obtains Acc score of 84.94% and AUROC score of 80.62% which are higher than the Acc score of 84.79% and AUROC score of 74.86% on the LS-100 under the setting of 5-way 1-shot, respectively. Whether the PGOC and PAPN are used or not, the scores of Acc and AUROC of our method are improved by the PGFC.

In addition, both Acc and AUROC scores obtained by our method with PGOC are higher than the counterparts obtained by our method without PGOC on three datasets under two way-shot settings. For instance, it can be observed from the results of cases ⑤ and ⑦ that our method with PGOC obtains Acc score of 85.10% and AUROC score of 82.72% which are higher than the Acc score of 83.97% and AUROC score of 71.41% on the LS-100 under the setting of 5-way 1-shot, respectively. Whether the PGFC and PAPN are used or not, the scores of Acc and AUROC are improved by the PGOC.

Similarly, both Acc and AUROC scores obtained by our method with PAPN are higher than the counterparts obtained by our method without PAPN on three datasets under two way-shot settings. For example, we can see from the results of cases ① and ② that our method with PAPN obtains Acc score of 84.79% and AUROC score of 74.86% which are higher than the Acc score of 80.11% and AUROC score of 70.05% on the LS-100 under the setting of 5-way 1-shot, respectively. Whether the PGFC and PGOC are used or not, the scores of Acc and AUROC are improved by the PAPN.

By comparing the results of case ① with case ②, case ① with case ③, and case ① with case ⑤, it can be observed that when only one of PGFC, PGOC, or PAPN is added, the addition of PGOC alone results in the greatest improvement in the Acc and AUROC scores across the three datasets in most cases. Similarly, by comparing the results of case ⑧ with case ④, case ⑧ with case ⑥, and case ⑧ with case ⑦, it is found that when only one of PGFC, PGOC, or PAPN is removed, the removal of PGOC alone leads to the largest decrease in the Acc and AUROC scores across the three datasets in most cases. Among the three parts, the PGOC has the most significant contribution to the performance of our method.

Based on the above analyses, it can be concluded that Acc and AUROC scores are improved by the PGFC, PGOC, and PAPN. When all of them are used, our method obtains the highest Acc and AUROC scores. Hence, these three parts are complementary for improving the performance of our method.

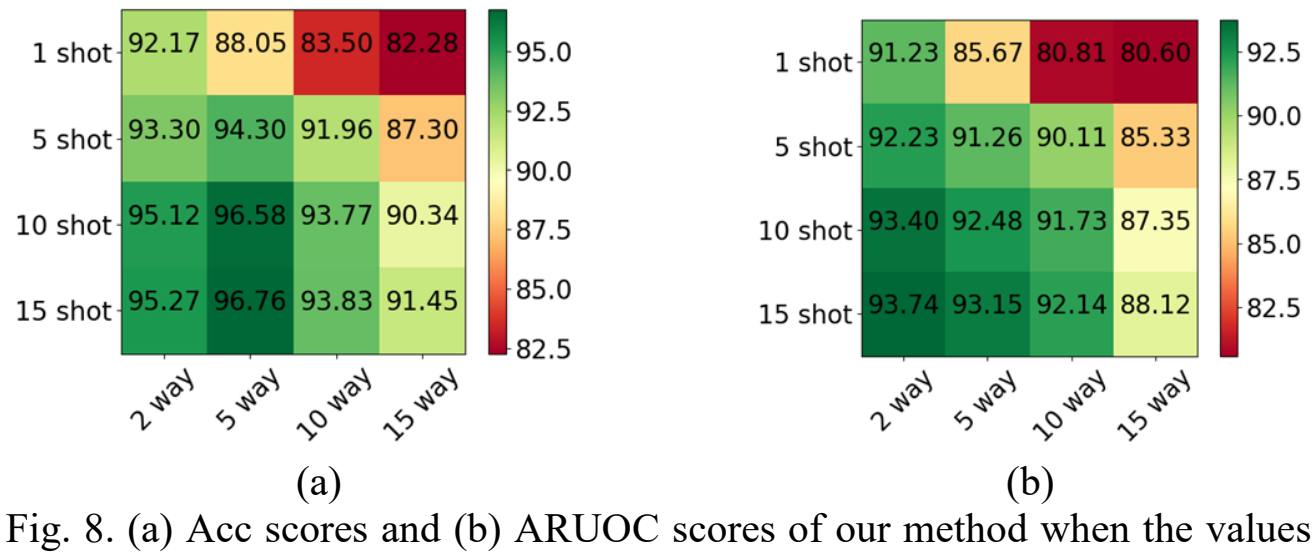

Fig. 8. (a) Acc scores and (b) ARUOC scores of our method when the values of $N$ are set to 2, 5, 10 or 15 and the values of $K$ are set to 1, 5, 10 or 15.

### H. *Extended Experiments*

To show the performance of our method from more aspects, four extended experiments are done. The NSynth-100 and a dataset of domestic environments [23] are adopted in the first three and the fourth extended experiments, respectively.

In the first extended experiment, we discuss the impacts of various settings of way-shot ($N$ and $K$) on the Acc and AUROC scores of our method. The values of $N$ are set to 2, 5, 10 or 15, while the values of $K$ are set to 1, 5, 10 or 15. Fig. 8 (a) and (b) shows the Acc and AUROC scores, respectively. The settings of way-shot have impacts on the Acc and AUROC scores.

The influences of way-shot settings have two characteristics. First, if we fix $N$ and increase $K$, the Acc and AUROC scores consistently increase. Main reason for this result is described as follows. When we increase $K$, we can adopt more support samples to generate the prototypes of few-shot classes. As a result, the generated prototypes can effectively represent the few-shot and open-set classes, and thus higher Acc and AUROC scores can be obtained by our method. When we increase $K$ from 1 to 5, the maximum Acc and AUROC gains are 8.46% (91.96% - 83.50%) and 10.30% (90.11% - 80.81%), respectively. The above Acc and AUROC gains are larger than the maximum Acc gain of 4.15% (91.45% - 87.30%) and the maximum AUROC gain of 2.79% (88.12% - 85.33%) when we increase $K$ from 5 to 15. In other words, when we increase the number of support samples of each few-shot class from 1 to 5, a significant gain of Acc and AUROC can be obtained. When we increase the number of support samples of each few-shot class from 5 to 15, the Acc and AUROC gains are no longer significant. Main reason for this result is described as follows. It is difficult to effectively represent the characteristics of the class using one support sample. However, it is not difficult to effectively characterize the characteristics of the class using five support samples. When we increase the number of support samples from 5 to 15, the characteristics represented by each sample may have redundancy, and thus the gains in Acc and AUROC become no longer significant.

Second, if we fix $K$ and increase $N$, the Acc and AUROC scores stably decrease. Main reason for this result is as follows. When the number of novel classes raises, confusions between various novel classes increase. The difficulty of distinguishing few-shot classes from open-set classes, and the difficulty of recognizing few-shot classes increase. Hence, the Acc and AUROC scores decrease.

In the second extended experiment, we discuss the impact of Openness [66], [67] on the performance of our method. The Openness measures the proportion of open-set classes in all

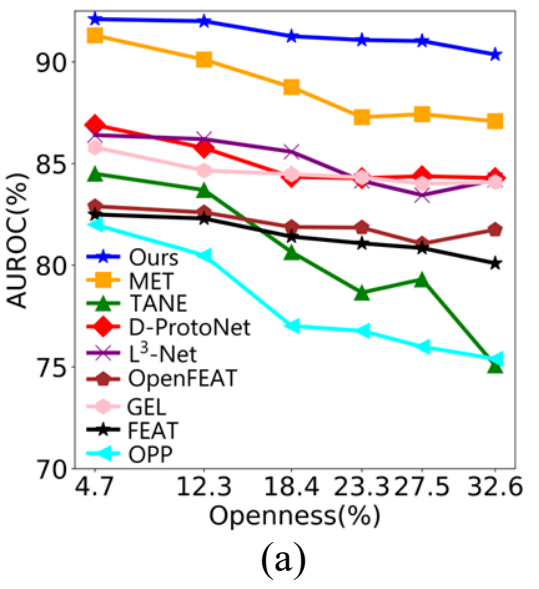

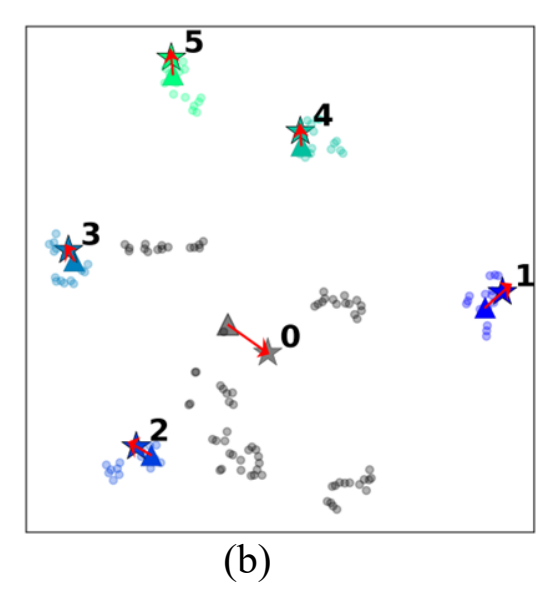

Fig. 9. (a) ARUOC scores against various Openness values. (b) Visualization of query embeddings and prototypes. The black and other colored dots denote the query embeddings of open-set and few-shot classes, respectively. The pentagrams and triangles denote prototypes that are generated with and without the proposed classifier, respectively. Red arrows denote movement of prototypes caused by the proposed classifier. Digit 0 denotes one open-set class, while digits 1 to 5 represent five few-shot classes.

novel classes of a query set during each test, and is defined by

$$Openness{=}1\text{-}\sqrt{\frac{2N_k}{2N_k+N_u}}, \quad (16)$$

where $N_k$ and $N_u$ denote the number of few-shot and open-set classes in a query set of each test, respectively. We fix $N_k$ and adjust $N_u$ to obtain different values of Openness. For each $N_u$ value, we randomly select query samples of ($N_u$+$N_k$) classes from the NSynth-100. We repeat the evaluation 600 times and take the average as the final result for each $N_u$ value. In previous work, the number of few-shot classes is typically set to 5, and the number of open-set classes is not more than twice the number of few-shot classes. Hence, $N_k$ is set to 5, while $N_u$ is set to 1, 3, 5, 7, 9, and 12. Fig. 9 (a) shows the ARUOC scores against various Openness values. When the Openness increases from 4.7% ($N_u$=1) to 32.6% ($N_u$=12), the AUROC score decrease of our method is slow. Our method is insensitive to changes in Openness. In addition, our method consistently obtains the highest AUROC scores among all methods.

In the third extended experiment, we depict the distributions of query embeddings and prototypes that are generated with and without the proposed classifier (consisting of the PGFC and PGOC). We use t-SNE [68] to map query embeddings and prototypes into two-dimensional space as shown in Fig. 9 (b). We use the Python libraries of *scikit-learn* and *matplotlib* to reduce the dimension of query embeddings and prototypes and to plot Fig. 9 (b), respectively. Five few-shot classes and their open-set class are randomly selected from the NSynth-100. Fig. 9 (b) shows that the prototypes (marked by pentagrams) generated with the proposed classifier move away from the confusion areas. That is, the intervals between the prototypes generated with the proposed classifier are larger than those between the prototypes (marked by triangles) generated without the proposed classifier. Hence, the proposed classifier can improve the representational ability of the prototypes and obtain more discriminative decision boundaries.

In the fourth extended experiment, we compare our method with previous methods on a dataset of domestic environments released in [23] for strengthening our method's advantage. The dataset consists of 1360 clips from 34 classes (including 24 types of pattern-sounds and 10 types of unwanted-sounds). Each class has 40 audio clips. In the pre-training step, the 24 types of pattern-sounds are used as the base classes to pre-train the encoder. In the meta-training step, few-shot learning (episodic learning of $N$-way $K$-shot) are performed using the

TABLE XII
RESULTS OBTAINED BY DIFFERENT METHODS ON A DATASET OF DOMESTIC ENVIRONMENTS (IN %)

| Methods | Dataset of domestic environments | | | |
|---|---|---|---|---|
| | 5-way 1-shot | | 5-way 5-shot | |
| | Acc | AUROC | Acc | AUROC |
| FEAT [37] | 77.53 | 64.62 | 87.31 | 65.81 |
| $L^3$-Net [23] | 73.42 | 69.15 | 77.49 | 70.35 |
| D-ProtoNet [21] | 70.47 | 69.11 | 80.62 | 71.44 |
| OpenFEAT [22] | 66.13 | 53.20 | 70.88 | 53.29 |
| TANE [36] | 73.24 | 69.38 | 74.00 | 69.98 |
| GEL [38] | 76.20 | 58.15 | 74.29 | 68.71 |
| OPP [39] | 77.79 | 64.04 | 85.40 | 65.57 |
| MET [40] | 70.25 | 55.15 | 84.76 | 56.24 |
| Ours | **79.51** | **70.33** | **87.32** | **71.58** |

24 types of pattern-sounds. In each episodic learning, some pattern-sounds are used as few-shot classes and other pattern-sounds are used as open-set classes. In the testing session, some unwanted-sounds are used as few-shot classes and other unwanted-sounds are used as open-set classes for model testing. The way-shot settings (5-way 1-shot and 5-way 5-shot) and the previous methods are the same as those in Table VII. Table XII lists the Acc and AUROC scores of various methods on the dataset of domestic environments. Our method achieves the highest Acc and AUROC scores among all methods. This result indicates that our method also exceeds other methods on the dataset of domestic environments. In short, the Acc and AUROC scores obtained by our method possess generalizability on different datasets, and are higher than the counterparts obtained by other methods.

## V. CONCLUSIONS

In this work, we explored a new problem of few-shot open-set audio classification, and tried to address it using attention information-fused prototypes. To enable the prototypes of open-set classes to acquire rich information from base classes, we generated prototypes of augmented base-classes that were involved in the prototype's generation of open-set classes. A classifier is designed to make the prototypes of novel classes aware of the class-discriminative information of support and query embeddings and salient information of support embeddings. We effectively integrated several established techniques (e.g., prototype learning, attention mechanisms, and open-set modeling) to design a novel method framework according to the specificity of the FOAC problem, rather than simply combining them. Hence, the usage of the established techniques did not affect the novelty of our method.

Based on the method description and experimental results in the above sections, we can draw four conclusions. First, each main part of our method contributes to the increase of accuracy. Second, our method has advantage over prior methods in accuracy. Moreover, this advantage is of statistical significance for almost all prior methods. Third, compared to most prior methods, our method has advantage in computational load and disadvantage in memory overhead. Fourth, our method exceeds prior methods in accuracy across datasets.

Although our method achieved more satisfactory results than prior methods, it has two limitations. First, its complexity is so high that it cannot be directly deployed on terminals with limited resources. To deploy it on audio terminals, we will consider lightweight model design, such as distilling the trained model into a small one, compressing model parameters. Second, our method exhibits performance degradation crossing datasets. We will consider domain adaptation or generalization to overcome the impact of domain-shift on our method.

## REFERENCES

[1] K. Drossos, et al., "Sound event detection with depthwise separable and dilated convolutions," in *Proc. of IJCNN*, 2020, pp. 1-7.

[2] Z. Lin, et al., "Domestic activities clustering from audio recordings using convolutional capsule autoencoder network," in *Proc. of ICASSP*, 2021, pp. 835-839.

[3] Y. Li, et al., "Acoustic event diarization in TV/movie audios using deep embedding and integer linear programming," *Multimedia Tools and Applications*, vol. 78, no. 23, pp. 33999-34025, 2019.

[4] Y. Li, et al., "Using multi-stream hierarchical deep neural network to extract deep audio feature for acoustic events detection," *Multimedia Tools and Applications*, vol. 77, no. 1, pp. 897-916, 2018.

[5] H.K. Chon, et al., "Acoustic scene classification using aggregation of two-scale deep embeddings," in *Proc. of IEEE ICCT*, 2021, vol.4, pp.1341-1345.

[6] Y. Li, et al., "Acoustic scene classification using deep audio feature and BLSTM network," in *Proc. of ICALIP*, 2018, pp. 371-374.

[7] K. Wang, et al., "Low-complexity acoustic scene classification using deep space separable distillation module and multi-label learning," in *Technical Report of DCASE2023 Challenge*, 2023, pp. 1-4.

[8] Y. Li, et al., "Low-complexity acoustic scene classification using parallel attention-convolution network", in *Proc. of Interspeech*, 2024, pp. 1-5.

[9] W. Pang, et al., "Violence detection in videos based on fusing visual and audio information," in *Proc. of ICASSP*, 2021, pp. 2260-2264.

[10] W. Pang, Q. He, and Y. Li, "Predicting skeleton trajectories using a Skeleton-Transformer for video anomaly detection," *Multimedia Systems*, 2022, vol. 28, pp. 1481-1494.

[11] Y. Li, et al., "Speaker clustering by co-optimizing deep representation learning and cluster estimation," *IEEE TMM*, vol. 23, pp. 3377-3387, 2021.

[12] R. Tao, et al., "Audio-visual target speaker extraction with selective auditory attention," *IEEE TASLP*, vol. 33, pp. 797-811, 2025.

[13] Y. Li, et al., "Lightweight speaker verification using transformation module with feature grouping and fusion," *IEEE/ACM TASLP*, vol. 32, pp. 794-806, 2024.

[14] Y. Li, et al., "Speaker verification using attentive multi-scale convolutional recurrent network," *Applied Soft Computing*, 2022, vol. 126, Article no. 109291, pp. 1-11.

[15] M. Yang, et al., "Cross-domain few-shot open-set keyword spotting using keyword adaptation and prototype reprojection," in *Proc. of ICASSP*, 2025, pp. 1-5.

[16] Y. Chen, et al., "Deep enhancement spotting network for low-complexity keyword spotting in noisy environments," in *Proc. of ICASSP*, 2025, pp. 1-5.

[17] J. Zhan, et al., "A stage match for query-by-example spoken term detection based on structure information of query," in *Proc. of ICASSP*, 2021, pp. 6833-6837.

[18] X. Lan, et al., "A novel re-weighted CTC loss for data imbalance in speech keyword spotting," *Chinese J. of Electronics*, vol.32, no.3, pp.465-473, 2023.

[19] L.Y. Li, et al., "Few-shot open-set keyword spotting with multi-stage training," in *Proc. of APSIPA ASC*, 2024, pp. 1-5.

[20] M. Rusci, et al., "Few-shot open-set learning for on-device customization of keyword spotting systems," in *Proc. of Interspeech*, 2024, pp. 2768-2772.

[21] B. Kim, et al., "Dummy prototypical networks for few-shot open-set keyword spotting," in *Proc. of Interspeech*, 2022, pp. 4621- 4625.

[22] K.C. Kishan et al., "OpenFEAT: Improving speaker identification by open-set few-shot embedding adaptation with Transformer," in *Proc. of ICASSP*, 2022, pp. 7062-7066.

[23] J. Naranjo-Alcazar, et al. "An open-set recognition and few-shot learning dataset for audio event classification in domestic environments," *Pattern Recognition Letters*, vol. 164, pp. 40-45, 2022.

[24] T. Hospedales, et al., "Meta-learning in neural networks: A survey," *IEEE TPAMI*, vol. 44, no. 9, pp. 5149-5169, 2022.

[25] W. Sun, et al., "MetaModulation: Learning variational feature hierarchies for few-shot learning with fewer tasks," in *Proc. of ICML*, 2023, pp. 32847-32858.

[26] J. Snell, et al., "Prototypical networks for few-shot learning," in *Proc. 31st Int. Conf. Neural Inf. Process. Syst.*, 2017, pp. 4080-4090.

[27] N. Mishra, et al., "A simple neural attentive meta-learner," in *Proc. of ICLR*, 2018, pp. 1-8.

[28] C. Finn, et al., "Model-agnostic meta-learning for fast adaptation of deep networks," in *Proc. of ICML*, 2017, vol. 70, pp. 1126-1135.
[29] K. Li, et al., "Learning to optimize," in *Proc. of ICLR*, 2017, pp. 1-13.
[30] L. Wang, et al., "Improving generalization of meta-learning with inverted regularization at inner-level," in *Proc. of CVPR*, 2023, pp. 7826-7835.
[31] S. Kang, et al., "Meta-learning with a geometry-adaptive preconditioner," in *Proc. of CVPR*, 2023, pp. 16080-16090.
[32] K.L. Pavasovic, et al., "MARS: Meta-learning as score matching in the function space," in *Proc. of ICLR*, 2023, pp. 1-26.
[33] T. Nguyen, et al., "Transformer neural processes: Uncertainty-aware meta learning via sequence modeling," in *Proc. of ICML*, 2022, pp. 16569-16594.
[34] X. Luo, et al., "A closer look at few-shot classification again," in *Proc. of ICML*, 2023, pp. 23103-23123.
[35] H. Chen, et al., "Using context-guided data augmentation, lightweight CNN, and proximity detection techniques to improve site safety monitoring under occlusion conditions," *Safety science*, 2023, vol. 158, article no.: 105958.
[36] S. Huang, et al., "Task-adaptive negative envision for few-shot open-set recognition," in *Proc. of IEEE/CVF CVPR*, 2022, pp. 7161-7170.
[37] H.-J. Ye, et al., "Few-shot learning via embedding adaptation with set-to-set functions," in *Proc. of IEEE/CVF CVPR*, 2020, pp. 8805-8814.
[38] H. Wang, et al., "Glocal energy-based learning for few-shot open-set recognition," in *Proc. of IEEE/CVF CVPR*, 2023, pp. 7507-7516.
[39] L.-Y., Sun, et al., "Overall positive prototype for few-shot open-set recognition," *Pattern Recognition*, vol.151, pp.1-10, 2024.
[40] H. Sapkota, et al., "Meta evidential transformer for few-shot open-set recognition," in *Proc. of ICML*, 2024, pp. 43389-43406.
[41] S.V. Dibbo, et al., "LCANets++: Robust audio classification using multi-layer neural networks with lateral competition," in *Proc. of ICASSPW*, 2024, pp. 129-133.
[42] J. Parekh, et al., "Tackling interpretability in audio classification networks with non-negative matrix factorization," *IEEE/ACM TASLP*, vol. 32, pp. 1392-1405, 2024.
[43] B. He, et al., "Light gated multi mini-patch extractor for audio classification," in *Proc. of ICASSPW*, 2024, pp. 765-769.
[44] G. Sharma, et al., "Time-frequency scattergrams for biomedical audio signal representation and classification," *IEEE/ACM TASLP*, vol. 32, pp. 564-576, 2024.
[45] C. Papaioannou, et al., "LC-Protonets: Multi-label few-shot learning for world music audio tagging," *IEEE Open Journal of Signal Processing*, vol. 6, pp. 138-146, 2025.
[46] C. Heggan, et al., "On the transferability of large-scale self-supervision to few-shot audio classification," in *Proc. of ICASSPW*, 2024, pp. 515-519.
[47] C. Heggan, et al., "From pixels to waveforms: evaluating pre-trained image models for few-shot audio classification," in *Proc. of IJCNN*, 2024, pp. 1-8.
[48] X. Zhuang, et al., "Episodic fine-tuning prototypical networks for optimization-based few-shot learning: application to audio classification," in *Proc. of IEEE MLSP*, 2024, pp. 1-6.
[49] H.J. Lee, et al., "Deep generative replay with denoising diffusion probabilistic models for continual learning in audio classification," *IEEE Access*, vol. 12, pp. 134714-134727, 2024.
[50] W. Pian, et al., "Audio-visual class-incremental learning," in *Proc. of IEEE/CVF ICCV*, 2023, pp. 7765-7777.
[51] Y. Xiao, et al., "UCIL: An unsupervised class incremental learning approach for sound event detection," in *Proc. of ICASSP*, 2025, pp. 1-5.
[52] Y. Li, et al., "Few-shot class-incremental audio classification using stochastic classifier," in *Proc. of Interspeech*, 2023, pp. 4174-4178.
[53] Y. Li, et al., "Few-shot class-incremental audio classification using pseudo-incrementally trained embedding learner and continually updated stochastic classifier," *IEEE TASLP*, vol. 33, pp. 3880-3895, 2025.
[54] Y. Li, et al., "Few-shot class-incremental audio classification using dynamically expanded classifier with self-attention modified prototypes," *IEEE TMM*, vol. 26, pp. 1346-1360, 2024.
[55] W. Xie, et al., "Few-shot class-incremental audio classification via discriminative prototype learning," *Expert Systems with Applications*, vol. 225, 120044, pp. 1-13, 2023.
[56] W. Xie, et al., "Few-shot class-incremental audio classification using adaptively-refined prototypes," in *Proc. of Interspeech*, 2023, pp.301-305.
[57] Y. Li, et al., "Few-shot class-incremental audio classification with adaptive mitigation of forgetting and overfitting," *IEEE/ACM TASLP*, vol. 32, pp. 2297-2311, 2024.
[58] Y. Si, et al., "Fully few-shot class-incremental audio classification using expandable dual-embedding extractor," *in Proc. of Interspeech*, 2024, pp. 4788-4792.
[59] Y. Si, et al., "Fully few-shot class-incremental audio classification with adaptive improvement of stability and plasticity," *IEEE TASLP*, vol. 33, pp. 418-433, 2025.
[60] Y. Li, et al., "Acoustic scene clustering using joint optimization of deep embedding learning and clustering iteration," *IEEE TMM*, vol. 22, no. 6, pp. 1385-1394, 2020.
[61] W. Xie, et al., "Deep mutual attention network for acoustic scene classification," *Digital Signal Processing*, vol. 123, 2022, Art. no. 103450.
[62] Y. Li, et al., "Domestic activity clustering from audio via depthwise separable convolutional autoencoder network," in *Proc. MMSP*, 2022, pp. 1-6.
[63] K. He, et al., "Deep residual learning for image recognition," in *Proc. of IEEE/CVF CVPR*, 2016, pp. 770-778.
[64] A. Vaswani et al., "Attention is all you need," in *Proc. of NIPS*, 2017, pp. 5998-6008.
[65] J. Demšar, "Statistical comparisons of classifiers over multiple data sets," *Journal of Machine Learning Research*, vol.7, pp. 1-30, 2006.
[66] C. Geng, et al., "Recent advances in open set recognition: A survey," *IEEE TPAMI*, vol. 43, no. 10, pp. 3614-3631, 2021.
[67] W. J. Scheirer, et al., "Toward open set recognition," *IEEE TPAMI*, vol. 35, no. 7, pp. 1757-1772, 2013.
[68] L.v.d. Maaten, et al., "Visualizing data using t-SNE," *Journal of Machine Learning Research*, vol. 9, no. 96, pp. 2579-2605, 2008.